\documentclass[twocolumn,aps,prc,superscriptaddress,showpacs,floatfix]{revtex4}

\usepackage{epsfig}
\usepackage{amsmath}
\usepackage{multirow}
\usepackage{graphicx}
\usepackage{amsmath}
\usepackage{feynmf}
\usepackage[normalem]{ulem}  
\usepackage[dvips]{color} 

\renewcommand\sout{\bgroup \color{red} \ULdepth=-.5ex \ULset}

\newcommand{\Ex}[2]{\ifmmode{#1\times10^{#2}}\else{$#1\times10^{#2}$}\fi}

\begin{document}
\title{Heptaquarks with two heavy antiquarks in a simple chromomagnetic model}

\author{Aaron Park}
\email{aaron.park@yonsei.ac.kr}\affiliation{Department of Physics and Institute of Physics and Applied Physics, Yonsei University, Seoul 03722, Korea}
\author{Woosung Park}\email{diracdelta@yonsei.ac.kr}\affiliation{Department of Physics and Institute of Physics and Applied Physics, Yonsei University, Seoul 03722, Korea}
\author{Su Houng~Lee}\email{suhoung@yonsei.ac.kr}\affiliation{Department of Physics and Institute of Physics and Applied Physics, Yonsei University, Seoul 03722, Korea}
\date{\today}
\begin{abstract}
We investigate the symmetry property and the stability of the heptaquark containing two identical heavy antiquarks using color-spin interaction. We construct the wave function of the heptaquark from the Pauli exclusion principle in the  SU(3) breaking case. The stability of the heptaquark against the strong decay into one baryon and two mesons is discussed in a simple chromomagnetic model. We find that $q^2 s^3 \bar{s}^2$ with $I=0,S=\frac{5}{2}$ is the most stable heptaquark configuration that could be probed by reconstructing the $\Lambda+\phi+\phi$ invariant mass.
\end{abstract}


\maketitle

\section{Introduction}

Multiquark hadrons made of more than three quarks became a theme of interest since Jaffe predicted their existence using the bag model~\cite{Jaffe:1976ig,Jaffe:1976ih,Jaffe:1976yi}.
Unfortunately, extensive experimental search ruled out the existence of a deeply bound H dibaryon, and the initial excitement about the finding of $\Theta^+(1540)$\cite{Nakano:2003qx} faded away as further experimental study could not confirm it~\cite{Aubert:2004bm,Abe:2004ws,Bai:2004gk,Gorelov:2004bu,Stenson:2004yz,Abt:2004tz,Longo:2004ana,Armstrong:2004gp,Pinkenburg:2004ux,Adamovich:2004yk}.

On the other hand, there is a renewed interest in the subject triggered by the discovery of the $X(3872)$ by the  Belle Collaboration  in $B^+ \rightarrow K^{\pm}\pi^+ \pi^+ J/\psi$ \cite{Choi:2003ue}, which  was subsequently confirmed by  several other experiments \cite{Acosta:2003zx,Abazov:2004kp,Aubert:2004ns}.
Also, in the dibaryon sector, a resonance structure was finally observed in the $I(J^P)=0(3^+)$ channel by the WASA-at-COSY collaboration\cite{Adlarson:2011bh,Clement:2016vnl}.
Furthermore, LHCb collaboration has recently observed hidden-charm pentaquark states in the $J/\psi p$ invariant mass spectrum in the $\Lambda^0_b \rightarrow J/\psi K^- p$ process \cite{Aaij:2015tga}.  Subsequently, these states were studied using many theoretical approaches, such as the QCD sum rules \cite{Chen:2015moa,Wang:2015epa,Azizi:2016dhy}, the molecular approach \cite{Chen:2015loa,He:2015cea,Huang:2015uda,Shimizu:2016rrd} and the quark model \cite{Yang:2015bmv,Park:2017jbn}.   These experimental findings led to the   interest in the study of multiquark hadron states containing heavy quarks.
In fact, recent lattice calculations show that the H dibaryon becomes bound in the massive pion cases \cite{Beane:2010hg,Inoue:2010es}.

Multiquark configurations with heavy quarks were studied before.  Silvestre-Brac and Leandri searched stable $q^6$, $q^5 Q$ and $q^4 Q Q'$ system in the framework of a pure chromomagnetic Hamiltonian \cite{SilvestreBrac:1992yg,Leandri:1993zg,Leandri:1995zm}.  Heavy pentaquarks with heavy quarks were studied in quark models with color spin \cite{Gignoux:1987cn,Lipkin:1987sk}  and flavor spin  interaction\cite{Stancu:1998sm,Genovese:1997tm}.
Tetraquark states with two heavy quarks were also found to be stable against strong decay if the heavy quark mass was taken to be sufficiently large~\cite{Carlson:1987hh}.

Within the constituent quark model, stable multiquark configurations  arise from a large attraction in the color-spin interaction when more light quarks can interact with each other in a compact configuration.  However, at the same time, bringing additional quarks into compact configuration will generate additional kinetic energy compared to having isolated hadrons.  If the additional quarks or antiquarks are heavy, the additional kinetic energy can be made small while keeping the enhanced color-spin interaction among light quarks large, which can be effectively understood as additional diquark correlation\cite{Lee:2009rt}, compared to separated hadrons.
To further probe configurations with heavy quarks in yet another multiquark configuration, we will consider heptaquarks with two heavy antiquarks.

Heptaquark composed of five quarks and two anti quarks has been studied by only a few researchers \cite{Bander:1994sp,Bicudo:2003rw,NunezV.:2004mw}.
Ref.\cite{Bander:1994sp} suggested that as long as there is a stable meson state composed of two heavy quarks and two light quarks, there will be a stable heptaquark state within the chiral soliton model.  In fact, within a constituent quark model, there will be a stable tetraquark state with two heavy quarks or antiquarks~\cite{Carlson:1987hh}.  Hence, such configurations are also part of the configurations to be probed in this work using a constituent quark model with color-spin interaction.

To study the possible existence of compact exotic hadrons,
one first has to inspect the configuration with the  most attractive color-spin interaction. For example, the color-spin interaction in a  H-dibaryon is more attractive than that from the two separate $\Lambda \Lambda$ system because the former allows for three most attractive diquark configuration while the later has two separated diquark configurations. The most attractive diquark configuration is the maximally antisymmetric configuration in terms of color $\otimes$ flavor $\otimes$ spin.
In terms of two quark configurations, there are four states satisfying the Pauli exclusion principle. We represent them together with the matrix elements for the invariant appearing in the color spin interaction in Table~\ref{two-quark-interaction}, which can be obtained from
\begin{flalign}
& -\sum_{i<j}^N \lambda_i \lambda_j \sigma_i \cdot \sigma_j  & \nonumber \\
& =  \bigg[ \frac{4}{3} N(N-6) +4 I(I+1)+ \frac{4}{3} S(S+1) +2 C_c \bigg],&
\end{flalign}
where $N$ is the total number of quarks, $C_c=\frac{1}{4} \lambda^2$ the color Casimir operator, $S$ the spin and $I$ the isospin of the system.  Using $C_c=\frac{4}{3},\frac{10}{3}$ for color anti-triplet and sextet respectively, one notes, as given in the  table, that while the most attractive channel has spin 0, the spin 1 state also has an attractive combination.
For two antiquarks, we can use the same table simply by replacing $\bar{\textbf{3}}$ and $\textbf{6}$ with $\textbf{3}$ and $\bar{\textbf{6}}$ for color and flavor state, respectively.
For the quark antiquark configuration, we can construct a similar table as given in the lower part of Table~\ref{two-quark-interaction}.   Typically, the color-spin interaction is inversely proportional to the two constituent quark masses involved $1/(m_im_j)$.  Hence, in forming a multiquark configurations,  if the addition involves a light quark and a light antiquark, the addition will just fall apart into a meson state.  On the other hand, if the addition is composed of a light quark and a heavy antiquark, it could become energetically favorable to be in a compact configuration.
To probe such a possibility systematically, we investigate the symmetry property and the stability of the heptaquark containing two identical heavy antiquarks in a simple chromomagnetic model.
\begin{table}
\caption{The classification of two quarks and quark-antiquark color-spin interaction, with $\lambda_i, \sigma_i$ respectively representing the color and spin matrix of the $i$ quark. Two quarks state is determined to satisfy Pauli exclusion principle. We denote antisymmetric and symmetric state as $A$ and $S$ respectively. In the parenthesis, multiplet state is represented.}
\begin{center}
\begin{tabular}{c|c|c|c|c}
\hline
\hline
& \multicolumn{4}{c}{$qq$} \\
\hline
Color & $A(\bar{3})$ & $S(6)$ & $A(\bar{3})$ & $S(6)$ \\
\hline
Flavor & $A(\bar{3})$ & $A(\bar{3})$ & $S(6)$ &$S(6)$ \\
\hline
Spin & $A(1)$ & $S(3)$ & $S(3)$ & $A(1)$ \\
\hline
$ -\lambda_i \lambda_j \sigma_i \cdot \sigma_j$ & $-8$ & $-\frac{4}{3}$ & $\frac{8}{3}$ & $4$ \\
\hline
\end{tabular}
\begin{tabular}{c|c|c|c|c}
\hline
& \multicolumn{4}{c}{$q \bar{q}$} \\
\hline
Color & $(1)$ & $(8)$ & $(1)$ & $(8)$ \\
\hline
Spin & $A(1)$ & $A(1)$ & $S(3)$ & $S(3)$ \\
\hline
$ -\lambda_i \lambda_j \sigma_i \cdot \sigma_j$ & $-16$ & $2$ & $\frac{16}{3}$ & $-\frac{2}{3}$ \\
\hline
\hline
\end{tabular}
\end{center}
\label{two-quark-interaction}
\end{table}

This paper is organized as follows. We first explain why kinetic energy favors
compact heptaquarks containing  heavy flavors in Sec.~\ref{why heavy heptaquark}.  In Sec.~\ref{Color and spin basis functions}, we represent the color and spin basis functions of the heptaquark configuration. In Sec.~\ref{Wave function}, we construct the flavor $\otimes$ color $\otimes$ spin part of the wave function of the heptaquark in order to satisfy the Pauli exclusion principle. In Sec.~\ref{Color spin interaction}, we represent the color-spin interaction part of the Hamiltonian. In Sec.~\ref{Results}, we calculate the binding potential of the heptaquark and plot the results as a function of the  light/heavy quark mass ratio parameter $\eta$. Finally, we summarize our results in Sec.~\ref{summary}.\\
\section{Why heavy heptaquark?}
\label{why heavy heptaquark}
In this work, we investigate the stability of the heptaquark using hyperfine potential. Even if the hyperfine potential is attractive for a given heptaquark configuration, it cannot form a compact stable state if the additional repulsion from kinetic energy is large. Hence, we need to consider which flavor state makes the additional kinetic energy lower. In the remaining part of this paper, we consider only hyperfine potential, but we can simply estimate the additional kinetic energy using the following coordinate system and simple Gaussian spatial function.

If we label the  five light quarks as ($i=1\sim5$) and the two heavy antiquarks as ($i=6,7$), then we can choose the Jacobi coordinate system as follows.
\begin{align}
  &\sum^7_{i=1} \frac{1}{2}m_i \dot{\textbf{r}}_i^2-\frac{1}{2}M \dot{\textbf{r}}_{CM}^2=\sum^6_{i=1} \frac{1}{2}M_i \dot{\textbf{x}}_i^2 \nonumber\\
  &\mathrm{where}~ M=\sum^7_{i=1} m_i,~M_1=M_2=m_u, \nonumber\\
  &M_3=M_4=\frac{2m_u m_Q}{m_u+m_Q},~ M_5=\frac{5m_u(m_u+m_Q)}{2(4m_u+m_Q)} \nonumber\\
  &M_6=\frac{7(m_u+m_Q)(4m_u+m_Q)}{10(5m_u+2m_Q)} \nonumber\\
  &\textbf{r}_{CM}=\frac{1}{M}\sum^7_{i=1} m_i \textbf{r}_i
\end{align}
\begin{align}
  \textbf{x}_1=&\frac{1}{\sqrt{2}}(\textbf{r}_1-\textbf{r}_2) \nonumber\\
  \textbf{x}_2=&\sqrt{\frac{2}{3}}(\frac{1}{2}\textbf{r}_1+\frac{1}{2}\textbf{r}_2-\textbf{r}_3) \nonumber\\
  \textbf{x}_3=&\frac{1}{\sqrt{2}}(\textbf{r}_4-\textbf{r}_6) \nonumber\\
  \textbf{x}_4=&\frac{1}{\sqrt{2}}(\textbf{r}_5-\textbf{r}_7) \nonumber\\
  \textbf{x}_5=&\sqrt{\frac{6}{5}} ( \frac{1}{3} \textbf{r}_1+ \frac{1}{3} \textbf{r}_2 +\frac{1}{3} \textbf{r}_3 - \frac{m_u}{m_u+m_Q}\textbf{r}_4-\frac{m_Q}{m_u+m_Q}\textbf{r}_6 ) \nonumber\\
  \textbf{x}_6=&\sqrt{\frac{10}{7}} \{ \frac{1}{4m_u+m_Q}(m_u \textbf{r}_1+m_u \textbf{r}_2+m_u \textbf{r}_3+m_u \textbf{r}_4 \nonumber\\
  &+m_Q \textbf{r}_6)-\frac{1}{m_u+m_Q}(m_u \textbf{r}_5+m_Q \textbf{r}_7)\}.
\end{align}

This coordinate system can describe the  decay mode of the heptaquark consisting of one baryon and two mesons. In this coordinate system, $\textbf{x}_1$, $\textbf{x}_2$ describe the baryon system, while  $\textbf{x}_3$, $\textbf{x}_4$ represent the relative quark distances for the two mesons, respectively. If we chose a simple Gaussian form as the spatial function, then we can calculate the kinetic energy of the heptaquark as follows.
\begin{align}
  R=e^{-a_1 \textbf{x}_1^2 -a_2 \textbf{x}_2^2 -a_3 \textbf{x}_3^2 -a_4 \textbf{x}_4^2 -a_5 \textbf{x}_5^2 -a_6 \textbf{x}_6^2},
\end{align}
\begin{align}
  T=\sum^6_{i=1}\frac{\textbf{p}_i^2}{2M_i}= \sum^6_{i=1} \frac{3\hbar^2}{2M_i}a_i.
\end{align}

From the above expression, we can find that the additional kinetic energy of the heptaquark is $\frac{3\hbar^2}{2M_5}a_5+\frac{3\hbar^2}{2M_6}a_6$, corresponding to the additional kinetic energy from bringing in the two meson type of quark antiquark pair into a compact configuration.   In the heavy quark limit, $M_6 \rightarrow \infty$ making one of the additional kinetic terms zero, hence there is no penalty in the kinetic energy, while the extra light quark might contribute attractively to the pentaquark configuration.
On the other hand, if we chose only one  antiquark to be heavy, then the reduced masses become $M_5=m_u$ and $M_6=\frac{7m_u(m_u+m_Q)}{2(6m_u+m_Q)}$, so that both of the  additional terms survive in the heavy quark mass limit.
Therefore, we can conclude that if we want to make the additional kinetic energy of the heptaquark sufficiently small, one needs to  include at least two heavy quarks.
However, one still needs to weigh in the attraction from the color spin interaction to determine which combination generates the most attractive heptaquark configuration, which is the subject of this work.
It should be noted that if we want to make both additional terms zero, then we have to replace one more quark with heavy quark; such configurations with three heavy quarks however is experimentally quite difficult to produce and will not be considered here.

\section{Color and spin basis functions}
\label{Color and spin basis functions}

\subsection{Color basis function}
We can represent the color state of the heptaquark using color decomposition in the $SU(3)$ fundamental representation as follows.
  \begin{align}
    (\textbf{3} \otimes \textbf{3} \otimes \textbf{3}) &\otimes (\textbf{3} \otimes \bar{\textbf{3}}) \otimes (\textbf{3} \otimes \bar{\textbf{3}}) \nonumber\\
    =& (\textbf{1} \oplus \textbf{8} \oplus \textbf{8} \oplus \textbf{10}) \otimes (\textbf{1} \oplus \textbf{8}) \otimes (\textbf{1} \oplus \textbf{8}) \nonumber\\
    =& (\textbf{1} \oplus \textbf{8} \oplus \textbf{8} \oplus \textbf{10}) \nonumber\\
    & \otimes (\textbf{1} \oplus \textbf{1} \oplus \textbf{8} \oplus \textbf{8} \oplus \textbf{8} \oplus \textbf{8} \oplus \textbf{10} \oplus \bar{\textbf{10}} \oplus \textbf{27})
  \end{align}
Among the above states, two $(\textbf{1} \otimes \textbf{1})$, eight $(\textbf{8} \otimes \textbf{8})$ and one $(\textbf{10} \otimes \bar{\textbf{10}})$ states can form the color singlet state. Therefore, there are eleven color basis functions for heptaquark.
However, there is a more efficient way to represent the color basis of heptaquark using the Young-Yamanouchi basis.

The color state of two antiquarks is a triplet or an anti-sextet. Therefore, in order to construct the color singlet heptaquark state, the color state of five quarks should be an anti-triplet or a sextet. For five quarks, the number of the  Young-Yamanouchi basis of anti-triplets and sextets are five and six, respectively. Therefore, we can represent eleven color basis functions for heptaquark as follows.

\begin{widetext}
\begin{align}
  &\vert C_1 \rangle= \left(
  \begin{tabular}{|c|c|}
    \hline
    1 & 2   \\
    \hline
    3 & 4  \\
    \hline
    5   \\
    \cline{1-1}
  \end{tabular},
  \begin{tabular}{|c|}
    \hline
    $\bar{6}$ \\
    \hline
    $\bar{7}$ \\
    \hline
  \end{tabular}
  \right),
  \vert C_2 \rangle= \left(
  \begin{tabular}{|c|c|}
    \hline
    1 & 3   \\
    \hline
    2 & 4  \\
    \hline
    5   \\
    \cline{1-1}
  \end{tabular},
  \begin{tabular}{|c|}
    \hline
    $\bar{6}$ \\
    \hline
    $\bar{7}$ \\
    \hline
  \end{tabular}
  \right),
  \vert C_3 \rangle= \left(
  \begin{tabular}{|c|c|}
    \hline
    1 & 2   \\
    \hline
    3 & 5  \\
    \hline
    4   \\
    \cline{1-1}
  \end{tabular},
  \begin{tabular}{|c|}
    \hline
    $\bar{6}$ \\
    \hline
    $\bar{7}$ \\
    \hline
  \end{tabular}
  \right),
  \vert C_4 \rangle= \left(
  \begin{tabular}{|c|c|}
    \hline
    1 & 3   \\
    \hline
    2 & 5  \\
    \hline
    4   \\
    \cline{1-1}
  \end{tabular},
  \begin{tabular}{|c|}
    \hline
    $\bar{6}$ \\
    \hline
    $\bar{7}$ \\
    \hline
  \end{tabular}
  \right),
  \vert C_5 \rangle= \left(
  \begin{tabular}{|c|c|}
    \hline
    1 & 4   \\
    \hline
    2 & 5  \\
    \hline
    3   \\
    \cline{1-1}
  \end{tabular},
  \begin{tabular}{|c|}
    \hline
    $\bar{6}$ \\
    \hline
    $\bar{7}$ \\
    \hline
  \end{tabular}
  \right),\nonumber \\
  &\vert C_6 \rangle= \left(
  \begin{tabular}{|c|c|c|}
    \hline
    1 & 2 & 3   \\
    \hline
    4 \\
    \cline{1-1}
    5 \\
    \cline{1-1}
  \end{tabular},
  \begin{tabular}{|c|c|}
    \hline
    $\bar{6}$ & $\bar{7}$ \\
    \hline
  \end{tabular}
  \right),
  \vert C_7 \rangle= \left(
  \begin{tabular}{|c|c|c|}
    \hline
    1 & 2 & 4   \\
    \hline
    3 \\
    \cline{1-1}
    5 \\
    \cline{1-1}
  \end{tabular},
  \begin{tabular}{|c|c|}
    \hline
    $\bar{6}$ & $\bar{7}$ \\
    \hline
  \end{tabular}
  \right),
  \vert C_8 \rangle= \left(
  \begin{tabular}{|c|c|c|}
    \hline
    1 & 3 & 4   \\
    \hline
    2 \\
    \cline{1-1}
    5 \\
    \cline{1-1}
  \end{tabular},
  \begin{tabular}{|c|c|}
    \hline
    $\bar{6}$ & $\bar{7}$ \\
    \hline
  \end{tabular}
  \right),
  \vert C_9 \rangle= \left(
  \begin{tabular}{|c|c|c|}
    \hline
    1 & 2 & 5   \\
    \hline
    3 \\
    \cline{1-1}
    4 \\
    \cline{1-1}
  \end{tabular},
  \begin{tabular}{|c|c|}
    \hline
    $\bar{6}$ & $\bar{7}$ \\
    \hline
  \end{tabular}
  \right), \nonumber\\
  &\vert C_{10} \rangle= \left(
  \begin{tabular}{|c|c|c|}
    \hline
    1 & 3 & 5   \\
    \hline
    2 \\
    \cline{1-1}
    4 \\
    \cline{1-1}
  \end{tabular},
  \begin{tabular}{|c|c|}
    \hline
    $\bar{6}$ & $\bar{7}$ \\
    \hline
  \end{tabular}
  \right),
  \vert C_{11} \rangle= \left(
  \begin{tabular}{|c|c|c|}
    \hline
    1 & 4 & 5   \\
    \hline
    2 \\
    \cline{1-1}
    3 \\
    \cline{1-1}
  \end{tabular},
  \begin{tabular}{|c|c|}
    \hline
    $\bar{6}$ & $\bar{7}$ \\
    \hline
  \end{tabular}
  \right).
\end{align}
\end{widetext}
In the appendix, we present the color basis of the heptaquark using the tensor notation.
The expectation value of all the color operators for the heptaquarks can be obtained  using this color basis.

\subsection{Spin basis function}
Seven quark system can have spin $\frac{7}{2},\frac{5}{2},\frac{3}{2}$ and $\frac{1}{2}$. We represent the spin basis functions in terms of Young-Yamanouchi basis for each spin values.
\begin{itemize}
  \item $S=\frac{7}{2}$ : one basis function with Young tableau [7] \\
  \begin{align}
    \vert S^{\frac{7}{2}}_{1} \rangle=
    \begin{tabular}{|c|c|c|c|c|c|c|}
      \hline
      1 & 2 & 3 & 4 & 5 & 6 & 7 \\
      \hline
    \end{tabular}
  \end{align}
  \item $S=\frac{5}{2}$ : six basis functions with Young tableau [6,1] \\
  \begin{align}
    \vert S^{\frac{5}{2}}_{1} \rangle=
    \begin{tabular}{|c|c|c|c|c|c|}
      \hline
      1 & 2 & 3 & 4 & 5 & 6  \\
      \hline
      7 \\
      \cline{1-1}
    \end{tabular},~
    \vert S^{\frac{5}{2}}_{2} \rangle=
    \begin{tabular}{|c|c|c|c|c|c|}
      \hline
      1 & 2 & 3 & 4 & 5 & 7  \\
      \hline
      6 \\
      \cline{1-1}
    \end{tabular}, \nonumber\\
    \vert S^{\frac{5}{2}}_{3} \rangle=
    \begin{tabular}{|c|c|c|c|c|c|}
      \hline
      1 & 2 & 3 & 4 & 6 & 7  \\
      \hline
      5 \\
      \cline{1-1}
    \end{tabular},~
    \vert S^{\frac{5}{2}}_{4} \rangle=
    \begin{tabular}{|c|c|c|c|c|c|}
      \hline
      1 & 2 & 3 & 5 & 6 & 7  \\
      \hline
      4 \\
      \cline{1-1}
    \end{tabular}, \nonumber\\
    \vert S^{\frac{5}{2}}_{5} \rangle=
    \begin{tabular}{|c|c|c|c|c|c|}
      \hline
      1 & 2 & 4 & 5 & 6 & 7  \\
      \hline
      3 \\
      \cline{1-1}
    \end{tabular},~
    \vert S^{\frac{5}{2}}_{6} \rangle=
    \begin{tabular}{|c|c|c|c|c|c|}
      \hline
      1 & 3 & 4 & 5 & 6 & 7  \\
      \hline
      2 \\
      \cline{1-1}
    \end{tabular}.
  \end{align}
  \item $S=\frac{3}{2}$ : fourteen basis functions with Young tableau [5,2] \\
  \begin{align}
    &\vert S^{\frac{3}{2}}_{1} \rangle=
    \begin{tabular}{|c|c|c|c|c|}
      \hline
      1 & 2 & 3 & 4 & 5 \\
      \hline
      6 & 7 \\
      \cline{1-2}
    \end{tabular},
    \vert S^{\frac{3}{2}}_{2} \rangle=
    \begin{tabular}{|c|c|c|c|c|}
      \hline
      1 & 2 & 3 & 4 & 6 \\
      \hline
      5 & 7 \\
      \cline{1-2}
    \end{tabular},
    \vert S^{\frac{3}{2}}_{3} \rangle=
    \begin{tabular}{|c|c|c|c|c|}
      \hline
      1 & 2 & 3 & 5 & 6 \\
      \hline
      4 & 7 \\
      \cline{1-2}
    \end{tabular}, \nonumber\\
    &\vert S^{\frac{3}{2}}_{4} \rangle=
    \begin{tabular}{|c|c|c|c|c|}
      \hline
      1 & 2 & 4 & 5 & 6 \\
      \hline
      3 & 7 \\
      \cline{1-2}
    \end{tabular},
    \vert S^{\frac{3}{2}}_{5} \rangle=
    \begin{tabular}{|c|c|c|c|c|}
      \hline
      1 & 3 & 4 & 5 & 6 \\
      \hline
      2 & 7 \\
      \cline{1-2}
    \end{tabular},
    \vert S^{\frac{3}{2}}_{6} \rangle=
    \begin{tabular}{|c|c|c|c|c|}
      \hline
      1 & 2 & 3 & 4 & 7 \\
      \hline
      5 & 6 \\
      \cline{1-2}
    \end{tabular}, \nonumber\\
    &\vert S^{\frac{3}{2}}_{7} \rangle=
    \begin{tabular}{|c|c|c|c|c|}
      \hline
      1 & 2 & 3 & 5 & 7 \\
      \hline
      4 & 6 \\
      \cline{1-2}
    \end{tabular},
    \vert S^{\frac{3}{2}}_{8} \rangle=
    \begin{tabular}{|c|c|c|c|c|}
      \hline
      1 & 2 & 4 & 5 & 7 \\
      \hline
      3 & 6 \\
      \cline{1-2}
    \end{tabular},
    \vert S^{\frac{3}{2}}_{9} \rangle=
    \begin{tabular}{|c|c|c|c|c|}
      \hline
      1 & 3 & 4 & 5 & 7 \\
      \hline
      2 & 6 \\
      \cline{1-2}
    \end{tabular}, \nonumber\\
    &\vert S^{\frac{3}{2}}_{10} \rangle=
    \begin{tabular}{|c|c|c|c|c|}
      \hline
      1 & 2 & 3 & 6 & 7 \\
      \hline
      4 & 5 \\
      \cline{1-2}
    \end{tabular},
    \vert S^{\frac{3}{2}}_{11} \rangle=
    \begin{tabular}{|c|c|c|c|c|}
      \hline
      1 & 2 & 4 & 6 & 7 \\
      \hline
      3 & 5 \\
      \cline{1-2}
    \end{tabular},
    \vert S^{\frac{3}{2}}_{12} \rangle=
    \begin{tabular}{|c|c|c|c|c|}
      \hline
      1 & 3 & 4 & 6 & 7 \\
      \hline
      2 & 5 \\
      \cline{1-2}
    \end{tabular}, \nonumber\\
    &\vert S^{\frac{3}{2}}_{13} \rangle=
    \begin{tabular}{|c|c|c|c|c|}
      \hline
      1 & 2 & 5 & 6 & 7 \\
      \hline
      3 & 4 \\
      \cline{1-2}
    \end{tabular},
    \vert S^{\frac{3}{2}}_{14} \rangle=
    \begin{tabular}{|c|c|c|c|c|}
      \hline
      1 & 3 & 5 & 6 & 7 \\
      \hline
      2 & 4 \\
      \cline{1-2}
    \end{tabular}.
  \end{align}
  \item $S=\frac{1}{2}$ : fourteen basis functions with Young tableau [4,3] \\
  \begin{align}
    &\vert S^{\frac{1}{2}}_{1} \rangle=
    \begin{tabular}{|c|c|c|c|}
      \hline
      1 & 2 & 3 & 4 \\
      \hline
      5 & 6 & 7 \\
      \cline{1-3}
    \end{tabular},
    \vert S^{\frac{1}{2}}_{2} \rangle=
    \begin{tabular}{|c|c|c|c|}
      \hline
      1 & 2 & 3 & 5 \\
      \hline
      4 & 6 & 7 \\
      \cline{1-3}
    \end{tabular},
    \vert S^{\frac{1}{2}}_{3} \rangle=
    \begin{tabular}{|c|c|c|c|}
      \hline
      1 & 2 & 4 & 5 \\
      \hline
      3 & 6 & 7 \\
      \cline{1-3}
    \end{tabular}, \nonumber\\
    &\vert S^{\frac{1}{2}}_{4} \rangle=
    \begin{tabular}{|c|c|c|c|}
      \hline
      1 & 3 & 4 & 5 \\
      \hline
      2 & 6 & 7 \\
      \cline{1-3}
    \end{tabular},
    \vert S^{\frac{1}{2}}_{5} \rangle=
    \begin{tabular}{|c|c|c|c|}
      \hline
      1 & 2 & 3 & 6 \\
      \hline
      4 & 5 & 7 \\
      \cline{1-3}
    \end{tabular},
    \vert S^{\frac{1}{2}}_{6} \rangle=
    \begin{tabular}{|c|c|c|c|}
      \hline
      1 & 2 & 4 & 6 \\
      \hline
      3 & 5 & 7 \\
      \cline{1-3}
    \end{tabular}, \nonumber\\
    &\vert S^{\frac{1}{2}}_{7} \rangle=
    \begin{tabular}{|c|c|c|c|}
      \hline
      1 & 3 & 4 & 6 \\
      \hline
      2 & 5 & 7 \\
      \cline{1-3}
    \end{tabular},
    \vert S^{\frac{1}{2}}_{8} \rangle=
    \begin{tabular}{|c|c|c|c|}
      \hline
      1 & 2 & 5 & 6 \\
      \hline
      3 & 4 & 7 \\
      \cline{1-3}
    \end{tabular},
    \vert S^{\frac{1}{2}}_{9} \rangle=
    \begin{tabular}{|c|c|c|c|}
      \hline
      1 & 3 & 5 & 6 \\
      \hline
      2 & 4 & 7 \\
      \cline{1-3}
    \end{tabular}, \nonumber\\
    &\vert S^{\frac{1}{2}}_{10} \rangle=
    \begin{tabular}{|c|c|c|c|}
      \hline
      1 & 2 & 3 & 7 \\
      \hline
      4 & 5 & 6 \\
      \cline{1-3}
    \end{tabular},
    \vert S^{\frac{1}{2}}_{11} \rangle=
    \begin{tabular}{|c|c|c|c|}
      \hline
      1 & 2 & 4 & 7 \\
      \hline
      3 & 5 & 6 \\
      \cline{1-3}
    \end{tabular},
    \vert S^{\frac{1}{2}}_{12} \rangle=
    \begin{tabular}{|c|c|c|c|}
      \hline
      1 & 3 & 4 & 7 \\
      \hline
      2 & 5 & 6 \\
      \cline{1-3}
    \end{tabular}, \nonumber\\
    &\vert S^{\frac{1}{2}}_{13} \rangle=
    \begin{tabular}{|c|c|c|c|}
      \hline
      1 & 2 & 5 & 7 \\
      \hline
      3 & 4 & 6 \\
      \cline{1-3}
    \end{tabular},
    \vert S^{\frac{1}{2}}_{14} \rangle=
    \begin{tabular}{|c|c|c|c|}
      \hline
      1 & 3 & 5 & 7 \\
      \hline
      2 & 4 & 6 \\
      \cline{1-3}
    \end{tabular}.
  \end{align}
\end{itemize}

\section{Wave function}
\label{Wave function}
There are two ways of constructing the wave function of the heptaquark. First, we can consider the flavor state of five quarks in SU(3) flavor symmetry.
In our previous work~\cite{Park:2016mez}, we have already classified all the possible flavor state for five light quarks. There are five possible flavor states for five quarks as follows:
\begin{eqnarray}
&&[\bar{3}]_F=
\begin{tabular}{|c|c|}
\hline
\quad \quad & \quad \quad  \\
\cline{1-2}
\quad \quad & \quad \quad  \\
\cline{1-2}
\quad \quad  \\
\cline{1-1}
\end{tabular}
,~
[6]_F=
\begin{tabular}{|c|c|c|}
\hline
\quad \quad & \quad \quad & \quad \quad \\
\hline
\multicolumn{1}{|c|}{\quad}  \\
\cline{1-1}
\quad \quad  \\
\cline{1-1}
\end{tabular}
,~
[\bar{15}]_F=
\begin{tabular}{|c|c|c|}
\hline
\quad \quad & \quad \quad & \quad \quad \\
\hline
\multicolumn{1}{|c|}{\quad} & \multicolumn{1}{|c|}{\quad}  \\
\cline{1-2}
\end{tabular}, \nonumber\\
&&[24]_F=
\begin{tabular}{|c|c|c|c|}
\hline
\quad \quad & \quad \quad & \quad \quad & \quad \quad \\
\hline
\multicolumn{1}{|c|}{\quad}  \\
\cline{1-1}
\end{tabular},
[21]_F=
\begin{tabular}{|c|c|c|c|c|}
\hline
\quad \quad & \quad \quad & \quad \quad & \quad \quad & \quad \quad \\
\hline
\end{tabular}.
\end{eqnarray}
For a given isospin and spin, we can choose the possible flavor states and construct the remaining part of the wave function using color and spin symmetry. In this description, the wave function should be antisymmetric for five light quarks and for two heavy antiquarks, respectively.

Second, we can calculate the wave function of the heptaquark in the SU(3) breaking case, fixing the position of strange quarks.
In the SU(3) flavor symmetry breaking case, we have to construct the
wave function to be antisymmetric separately for the  $u,d$ quarks, $s$ quarks and two heavy antiquarks, due to the Pauli exclusion principle.

Both approaches can be shown to give the same result~\cite{Park:2016cmg}. In this work, we follow the second method for convenience of calculation. To do this, we assume the spatial function to be symmetric such that the rest of the wave function represented by flavor $\otimes$ color $\otimes$ spin should be antisymmetric. Using color spin coupling scheme \cite{Park:2015nha,Park:2016mez}, we represent the wave function of the heptaquark by flavor $\otimes$ color-spin coupling basis.

\subsection{$q^5 \bar{Q}^2$ : \{12345\}\{67\}}
Here, we calculate the wave function of the heptaquark in the flavor $SU(3)$ breaking case and fix the position of each quarks on $q(1)q(2)q(3)q(4)q(5)\bar{Q}(6)\bar{Q}(7)$. In this case, the flavor $\otimes$ color $\otimes$ spin wave function should satisfy the following symmetry : \{12345\}\{67\}
\begin{align}
  \psi_{FCS}=\left(
  \begin{tabular}{|c|}
  \hline
  1 \\
  \hline
  2 \\
  \hline
  3 \\
  \hline
  4 \\
  \hline
  5 \\
  \hline
  \end{tabular},
  \begin{tabular}{|c|}
  \hline
  $\bar{6}$ \\
  \hline
  $\bar{7}$ \\
  \hline
  \end{tabular}\right)_{~FCS}
\end{align}
\begin{widetext}
\begin{itemize}
\item $I=\frac{5}{2}$
\begin{align}
  \psi_{FCS}=
  &\left(
  \begin{tabular}{|c|c|c|c|c|}
    \hline
    1 & 2 & 3 & 4 & 5 \\
    \hline
  \end{tabular},
  \begin{tabular}{|c|c|}
    \hline
    $\bar{6}$ & $\bar{7}$ \\
    \hline
  \end{tabular}\right)_F
  \otimes \left(
  \begin{tabular}{|c|}
    \hline
    1 \\
    \hline
    2 \\
    \hline
    3 \\
    \hline
    4 \\
    \hline
    5 \\
    \hline
  \end{tabular},
  \begin{tabular}{|c|}
    \hline
    $\bar{6}$ \\
    \hline
    $\bar{7}$ \\
    \hline
  \end{tabular}\right)_{CS}
  =F_1 \otimes CS_1
\end{align}
\item $I=\frac{3}{2}$
\begin{align}
  \psi_{FCS}
  &=\frac{1}{2}\bigg\{ \left(
  \begin{tabular}{|c|c|c|c|}
    \hline
    1 & 2 & 3 & 4  \\
    \hline
    5 \\
    \cline{1-1}
  \end{tabular},
  \begin{tabular}{|c|c|}
    \hline
    $\bar{6}$ & $\bar{7}$ \\
    \hline
  \end{tabular}\right)_F
  \otimes \left(
  \begin{tabular}{|c|c|}
    \hline
    1 & 5 \\
    \hline
    2 \\
    \cline{1-1}
    3 \\
    \cline{1-1}
    4 \\
    \cline{1-1}
  \end{tabular},
  \begin{tabular}{|c|}
    \hline
    $\bar{6}$ \\
    \hline
    $\bar{7}$ \\
    \hline
  \end{tabular}\right)_{CS}
  -\left(
  \begin{tabular}{|c|c|c|c|}
    \hline
    1 & 2 & 3 & 5  \\
    \hline
    4 \\
    \cline{1-1}
  \end{tabular},
  \begin{tabular}{|c|c|}
    \hline
    $\bar{6}$ & $\bar{7}$ \\
    \hline
  \end{tabular}\right)_F
  \otimes \left(
  \begin{tabular}{|c|c|}
    \hline
    1 & 4 \\
    \hline
    2 \\
    \cline{1-1}
    3 \\
    \cline{1-1}
    5 \\
    \cline{1-1}
  \end{tabular},
  \begin{tabular}{|c|}
    \hline
    $\bar{6}$ \\
    \hline
    $\bar{7}$ \\
    \hline
  \end{tabular}\right)_{CS} \nonumber \\
  &+\left(
  \begin{tabular}{|c|c|c|c|}
    \hline
    1 & 2 & 4 & 5  \\
    \hline
    3 \\
    \cline{1-1}
  \end{tabular},
  \begin{tabular}{|c|c|}
    \hline
    $\bar{6}$ & $\bar{7}$ \\
    \hline
  \end{tabular}\right)_F
  \otimes \left(
  \begin{tabular}{|c|c|}
    \hline
    1 & 3 \\
    \hline
    2 \\
    \cline{1-1}
    4 \\
    \cline{1-1}
    5 \\
    \cline{1-1}
  \end{tabular},
  \begin{tabular}{|c|}
    \hline
    $\bar{6}$ \\
    \hline
    $\bar{7}$ \\
    \hline
  \end{tabular}\right)_{CS}
  -\left(
  \begin{tabular}{|c|c|c|c|}
    \hline
    1 & 3 & 4 & 5  \\
    \hline
    2 \\
    \cline{1-1}
  \end{tabular},
  \begin{tabular}{|c|c|}
    \hline
    $\bar{6}$ & $\bar{7}$ \\
    \hline
  \end{tabular}\right)_F
  \otimes \left(
  \begin{tabular}{|c|c|}
    \hline
    1 & 2 \\
    \hline
    3 \\
    \cline{1-1}
    4 \\
    \cline{1-1}
    5 \\
    \cline{1-1}
  \end{tabular},
  \begin{tabular}{|c|}
    \hline
    $\bar{6}$ \\
    \hline
    $\bar{7}$ \\
    \hline
  \end{tabular}\right)_{CS}
  \bigg\} \nonumber \\
  &=\frac{1}{2}(F_1 \otimes CS_4 - F_2 \otimes CS_3 + F_3 \otimes CS_2 - F_4 \otimes CS_1)
\end{align}
\item $I=\frac{1}{2}$
\begin{align}
  \psi_{FCS}
  &=\frac{1}{\sqrt{5}} \bigg\{ \left(
  \begin{tabular}{|c|c|c|}
    \hline
    1 & 2 & 3  \\
    \hline
    4 & 5 \\
    \cline{1-2}
  \end{tabular},
  \begin{tabular}{|c|c|}
    \hline
    $\bar{6}$ & $\bar{7}$ \\
    \hline
  \end{tabular}\right)_F
  \otimes \left(
  \begin{tabular}{|c|c|}
    \hline
    1 & 4 \\
    \hline
    2 & 5 \\
    \hline
    3 \\
    \cline{1-1}
  \end{tabular},
  \begin{tabular}{|c|}
    \hline
    $\bar{6}$ \\
    \hline
    $\bar{7}$ \\
    \hline
  \end{tabular}\right)_{CS}
  -\left(
  \begin{tabular}{|c|c|c|}
    \hline
    1 & 2 & 4  \\
    \hline
    3 & 5 \\
    \cline{1-2}
  \end{tabular},
  \begin{tabular}{|c|c|}
    \hline
    $\bar{6}$ & $\bar{7}$ \\
    \hline
  \end{tabular}\right)_F
  \otimes \left(
  \begin{tabular}{|c|c|}
    \hline
    1 & 3 \\
    \hline
    2 & 5 \\
    \hline
    4 \\
    \cline{1-1}
  \end{tabular},
  \begin{tabular}{|c|}
    \hline
    $\bar{6}$ \\
    \hline
    $\bar{7}$ \\
    \hline
  \end{tabular}\right)_{CS}
  +\left(
  \begin{tabular}{|c|c|c|}
    \hline
    1 & 3 & 4  \\
    \hline
    2 & 5 \\
    \cline{1-2}
  \end{tabular},
  \begin{tabular}{|c|c|}
    \hline
    $\bar{6}$ & $\bar{7}$ \\
    \hline
  \end{tabular}\right)_F
  \otimes \left(
  \begin{tabular}{|c|c|}
    \hline
    1 & 2 \\
    \hline
    3 & 5 \\
    \hline
    4 \\
    \cline{1-1}
  \end{tabular},
  \begin{tabular}{|c|}
    \hline
    $\bar{6}$ \\
    \hline
    $\bar{7}$ \\
    \hline
  \end{tabular}\right)_{CS} \nonumber\\
  &+\left(
  \begin{tabular}{|c|c|c|}
    \hline
    1 & 2 & 5  \\
    \hline
    3 & 4 \\
    \cline{1-2}
  \end{tabular},
  \begin{tabular}{|c|c|}
    \hline
    $\bar{6}$ & $\bar{7}$ \\
    \hline
  \end{tabular}\right)_F
  \otimes \left(
  \begin{tabular}{|c|c|}
    \hline
    1 & 3 \\
    \hline
    2 & 4 \\
    \hline
    5 \\
    \cline{1-1}
  \end{tabular},
  \begin{tabular}{|c|}
    \hline
    $\bar{6}$ \\
    \hline
    $\bar{7}$ \\
    \hline
  \end{tabular}\right)_{CS}
  -\left(
  \begin{tabular}{|c|c|c|}
    \hline
    1 & 3 & 5  \\
    \hline
    2 & 4 \\
    \cline{1-2}
  \end{tabular},
  \begin{tabular}{|c|c|}
    \hline
    $\bar{6}$ & $\bar{7}$ \\
    \hline
  \end{tabular}\right)_F
  \otimes \left(
  \begin{tabular}{|c|c|}
    \hline
    1 & 2 \\
    \hline
    3 & 4 \\
    \hline
    5 \\
    \cline{1-1}
  \end{tabular},
  \begin{tabular}{|c|}
    \hline
    $\bar{6}$ \\
    \hline
    $\bar{7}$ \\
    \hline
  \end{tabular}\right)_{CS} \bigg\} \nonumber\\
  &=\frac{1}{\sqrt{5}}(F_1 \otimes CS_5 - F_2 \otimes CS_4 + F_3 \otimes CS_3 + F_4 \otimes CS_2 - F_5 \otimes CS_1)
\end{align}
\end{itemize}
\end{widetext}
\subsection{$q^4 s \bar{Q}^2$ : \{1234\}5\{67\}}
We fix the position of each quarks on $q(1)q(2)q(3)q(4)s(5)\bar{Q}(6)\bar{Q}(7)$. In this case, the flavor $\otimes$ color $\otimes$ spin wave function should satisfy the following symmetry : \{1234\}5\{67\} because there is no symmetry between $u$,$d$ quarks and $s$ quark in $SU(3)$ breaking case.
\begin{align}
  \psi_{FCS}=\left(
  \begin{tabular}{|c|}
    \hline
    1 \\
    \hline
    2 \\
    \hline
    3 \\
    \hline
    4 \\
    \hline
  \end{tabular},
  \begin{tabular}{|c|}
    \hline
    5 \\
    \hline
  \end{tabular},
  \begin{tabular}{|c|}
    \hline
    $\bar{6}$ \\
    \hline
    $\bar{7}$ \\
    \hline
  \end{tabular}\right)_{~FCS}
\end{align}
\begin{widetext}
  \begin{itemize}
    \item $I=2$
    \begin{align}
      \psi_{FCS}=
      &\left(
      \begin{tabular}{|c|c|c|c|}
        \hline
        1 & 2 & 3 & 4 \\
        \hline
      \end{tabular},
      \begin{tabular}{|c|}
        \hline
        5 \\
        \hline
      \end{tabular},
      \begin{tabular}{|c|c|}
        \hline
        $\bar{6}$ & $\bar{7}$ \\
        \hline
      \end{tabular}\right)_F
      \otimes \left(
      \begin{tabular}{|c|}
        \hline
        1 \\
        \hline
        2 \\
        \hline
        3 \\
        \hline
        4 \\
        \hline
      \end{tabular},
      \begin{tabular}{|c|}
        \hline
        5 \\
        \hline
      \end{tabular},
      \begin{tabular}{|c|}
        \hline
        $\bar{6}$ \\
        \hline
        $\bar{7}$ \\
        \hline
      \end{tabular}\right)_{CS}
      =F_1 \otimes CS_1
    \end{align}
    \item $I=1$
    \begin{align}
      \psi_{FCS}
      &=\frac{1}{\sqrt{3}} \bigg\{ \left(
      \begin{tabular}{|c|c|c|}
        \hline
        1 & 2 & 3  \\
        \hline
        4 \\
        \cline{1-1}
      \end{tabular},
      \begin{tabular}{|c|}
        \hline
        5 \\
        \hline
      \end{tabular},
      \begin{tabular}{|c|c|}
        \hline
        $\bar{6}$ & $\bar{7}$ \\
        \hline
      \end{tabular}\right)_F
      \otimes \left(
      \begin{tabular}{|c|c|}
        \hline
        1 & 4 \\
        \hline
        2 \\
        \cline{1-1}
        3 \\
        \cline{1-1}
      \end{tabular},
      \begin{tabular}{|c|}
        \hline
        5 \\
        \hline
      \end{tabular},
      \begin{tabular}{|c|}
        \hline
        $\bar{6}$ \\
        \hline
        $\bar{7}$ \\
        \hline
      \end{tabular}\right)_{CS}
      -\left(
      \begin{tabular}{|c|c|c|}
        \hline
        1 & 2 & 4  \\
        \hline
        3 \\
        \cline{1-1}
      \end{tabular},
      \begin{tabular}{|c|}
        \hline
        5 \\
        \hline
      \end{tabular},
      \begin{tabular}{|c|c|}
        \hline
        $\bar{6}$ & $\bar{7}$ \\
        \hline
      \end{tabular}\right)_F
      \otimes \left(
      \begin{tabular}{|c|c|}
        \hline
        1 & 3 \\
        \hline
        2 \\
        \cline{1-1}
        4 \\
        \cline{1-1}
      \end{tabular},
      \begin{tabular}{|c|}
        \hline
        5 \\
        \hline
      \end{tabular},
      \begin{tabular}{|c|}
        \hline
        $\bar{6}$ \\
        \hline
        $\bar{7}$ \\
        \hline
      \end{tabular}\right)_{CS} \nonumber\\
      &+\left(
      \begin{tabular}{|c|c|c|}
        \hline
        1 & 3 & 4  \\
        \hline
        2 \\
        \cline{1-1}
      \end{tabular},
      \begin{tabular}{|c|}
        \hline
        5 \\
        \hline
      \end{tabular},
      \begin{tabular}{|c|c|}
        \hline
        $\bar{6}$ & $\bar{7}$ \\
        \hline
      \end{tabular}\right)_F
      \otimes \left(
      \begin{tabular}{|c|c|}
        \hline
        1 & 2 \\
        \hline
        3 \\
        \cline{1-1}
        4 \\
        \cline{1-1}
      \end{tabular},
      \begin{tabular}{|c|}
        \hline
        5 \\
        \hline
      \end{tabular},
      \begin{tabular}{|c|}
        \hline
        $\bar{6}$ \\
        \hline
        $\bar{7}$ \\
        \hline
      \end{tabular}\right)_{CS} \bigg\} \nonumber\\
      &=\frac{1}{\sqrt{3}}(F_1 \otimes CS_3 - F_2 \otimes CS_2 + F_3 \otimes CS_1)
    \end{align}
    \item $I=0$
    \begin{align}
      \psi_{FCS}
      &=\frac{1}{\sqrt{2}} \bigg\{ \left(
      \begin{tabular}{|c|c|}
        \hline
        1 & 2 \\
        \hline
        3 & 4 \\
        \hline
      \end{tabular},
      \begin{tabular}{|c|}
        \hline
        5 \\
        \hline
      \end{tabular},
      \begin{tabular}{|c|c|}
        \hline
        $\bar{6}$ & $\bar{7}$ \\
        \hline
      \end{tabular} \right)_F
      \otimes
      \left(
      \begin{tabular}{|c|c|}
        \hline
        1 & 3 \\
        \hline
        2 & 4 \\
        \hline
      \end{tabular},
      \begin{tabular}{|c|}
        \hline
        5 \\
        \hline
      \end{tabular},
      \begin{tabular}{|c|}
        \hline
        $\bar{6}$ \\
        \hline
        $\bar{7}$ \\
        \hline
      \end{tabular}
      \right)_{CS}
      -\left(
      \begin{tabular}{|c|c|}
        \hline
        1 & 3 \\
        \hline
        2 & 4 \\
        \hline
      \end{tabular},
      \begin{tabular}{|c|}
        \hline
        5 \\
        \hline
      \end{tabular},
      \begin{tabular}{|c|c|}
        \hline
        $\bar{6}$ & $\bar{7}$ \\
        \hline
      \end{tabular} \right)_F
      \otimes
      \left(
      \begin{tabular}{|c|c|}
        \hline
        1 & 2 \\
        \hline
        3 & 4 \\
        \hline
      \end{tabular},
      \begin{tabular}{|c|}
        \hline
        5 \\
        \hline
      \end{tabular},
      \begin{tabular}{|c|}
        \hline
        $\bar{6}$ \\
        \hline
        $\bar{7}$ \\
        \hline
      \end{tabular}
      \right)_{CS}
      \bigg\} \nonumber\\
      &=\frac{1}{\sqrt{2}}(F_1 \otimes CS_2 - F_2 \otimes CS_1)
    \end{align}
  \end{itemize}
\end{widetext}
\subsection{$q^3 s^2 \bar{Q}^2$ : \{123\}\{45\}\{67\}}
We fix the position of each quarks on $q(1)q(2)q(3)s(4)s(5)\bar{Q}(6)\bar{Q}(7)$. In this case, the flavor $\otimes$ color $\otimes$ spin wave function should satisfy the following symmetry : \{123\}\{45\}\{67\}
\begin{align}
  \psi_{FCS}=\left(
  \begin{tabular}{|c|}
    \hline
    1 \\
    \hline
    2 \\
    \hline
    3 \\
    \hline
  \end{tabular},
  \begin{tabular}{|c|}
    \hline
    4 \\
    \hline
    5 \\
    \hline
  \end{tabular},
  \begin{tabular}{|c|}
    \hline
    $\bar{6}$ \\
    \hline
    $\bar{7}$ \\
    \hline
  \end{tabular}\right)_{~FCS}
\end{align}
\begin{widetext}
  \begin{itemize}
    \item $I=\frac{3}{2}$
    \begin{align}
      \psi_{FCS}=
      &\left(
      \begin{tabular}{|c|c|c|}
        \hline
        1 & 2 & 3 \\
        \hline
      \end{tabular},
      \begin{tabular}{|c|c|}
        \hline
        4 & 5 \\
        \hline
      \end{tabular},
      \begin{tabular}{|c|c|}
        \hline
        $\bar{6}$ & $\bar{7}$ \\
        \hline
      \end{tabular}\right)_F
      \otimes \left(
      \begin{tabular}{|c|}
        \hline
        1 \\
        \hline
        2 \\
        \hline
        3 \\
        \hline
      \end{tabular},
      \begin{tabular}{|c|}
        \hline
        4 \\
        \hline
        5 \\
        \hline
      \end{tabular},
      \begin{tabular}{|c|}
        \hline
        $\bar{6}$ \\
        \hline
        $\bar{7}$ \\
        \hline
      \end{tabular}\right)_{CS}
      =F_1 \otimes CS_1
    \end{align}
    \item $I=\frac{1}{2}$
    \begin{align}
      \psi_{FCS}
      &=\frac{1}{\sqrt{2}} \bigg\{ \left(
      \begin{tabular}{|c|c|}
        \hline
        1 & 2 \\
        \hline
        3 \\
        \cline{1-1}
      \end{tabular},
      \begin{tabular}{|c|c|}
        \hline
        4 & 5 \\
        \hline
      \end{tabular},
      \begin{tabular}{|c|c|}
        \hline
        $\bar{6}$ & $\bar{7}$ \\
        \hline
      \end{tabular} \right)_F
      \otimes \left(
      \begin{tabular}{|c|c|}
        \hline
        1 & 3 \\
        \hline
        2 \\
        \cline{1-1}
      \end{tabular},
      \begin{tabular}{|c|}
        \hline
        4 \\
        \hline
        5 \\
        \hline
      \end{tabular},
      \begin{tabular}{|c|}
        \hline
        $\bar{6}$ \\
        \hline
        $\bar{7}$ \\
        \hline
      \end{tabular} \right)_{CS}
      -\left(
      \begin{tabular}{|c|c|}
        \hline
        1 & 3 \\
        \hline
        2 \\
        \cline{1-1}
      \end{tabular},
      \begin{tabular}{|c|c|}
        \hline
        4 & 5 \\
        \hline
      \end{tabular},
      \begin{tabular}{|c|c|}
        \hline
        $\bar{6}$ & $\bar{7}$ \\
        \hline
      \end{tabular} \right)_F
      \otimes \left(
      \begin{tabular}{|c|c|}
        \hline
        1 & 2 \\
        \hline
        3 \\
        \cline{1-1}
      \end{tabular},
      \begin{tabular}{|c|}
        \hline
        4 \\
        \hline
        5 \\
        \hline
      \end{tabular},
      \begin{tabular}{|c|}
        \hline
        $\bar{6}$ \\
        \hline
        $\bar{7}$ \\
        \hline
      \end{tabular} \right)_{CS}
      \bigg\} \nonumber\\
      &=\frac{1}{\sqrt{2}}(F_1 \otimes CS_2 - F_2 \otimes CS_1)
    \end{align}
  \end{itemize}
\end{widetext}
\subsection{$s^3 q^2 \bar{Q}^2$ : \{123\}\{45\}\{67\}}
We fix the position of each quarks on $s(1)s(2)s(3)q(4)q(5)\bar{Q}(6)\bar{Q}(7)$ for convenience in calculation. In this case, the flavor $\otimes$ color $\otimes$ spin wave function should satisfy the following symmetry : \{123\}\{45\}\{67\}
\begin{align}
  \psi_{FCS}=\left(
  \begin{tabular}{|c|}
    \hline
    1 \\
    \hline
    2 \\
    \hline
    3 \\
    \hline
  \end{tabular},
  \begin{tabular}{|c|}
    \hline
    4 \\
    \hline
    5 \\
    \hline
  \end{tabular},
  \begin{tabular}{|c|}
    \hline
    $\bar{6}$ \\
    \hline
    $\bar{7}$ \\
    \hline
  \end{tabular}\right)_{~FCS}
\end{align}
  \begin{itemize}
    \item $I=1$ :
    The wave function is the same as in the case of $q^3 s^2 \bar{Q}^2$ with $I=\frac{3}{2}$.
    \item $I=0$
    \begin{align}
      \psi_{FCS}=
      &\left(
      \begin{tabular}{|c|c|c|}
        \hline
        1 & 2 & 3 \\
        \hline
      \end{tabular},
      \begin{tabular}{|c|}
        \hline
        4 \\
        \hline
        5 \\
        \hline
      \end{tabular},
      \begin{tabular}{|c|c|}
        \hline
        $\bar{6}$ & $\bar{7}$ \\
        \hline
      \end{tabular}\right)_F
      \otimes \left(
      \begin{tabular}{|c|}
        \hline
        1 \\
        \hline
        2 \\
        \hline
        3 \\
        \hline
      \end{tabular},
      \begin{tabular}{|c|c|}
        \hline
        4 & 5 \\
        \hline
      \end{tabular},
      \begin{tabular}{|c|}
        \hline
        $\bar{6}$ \\
        \hline
        $\bar{7}$ \\
        \hline
      \end{tabular}\right)_{CS} \nonumber\\
      &=F_1 \otimes CS_1
    \end{align}
  \end{itemize}
\subsection{$s^4 q \bar{Q}^2$ : \{1234\}5\{67\}}
We fix the position of each quarks on $s(1)s(2)s(3)s(4)q(5)\bar{Q}(6)\bar{Q}(7)$. In this case, the flavor $\otimes$ color $\otimes$ spin wave function should satisfy the following symmetry : \{1234\}5\{67\}
\begin{align}
  \psi_{FCS}=\left(
  \begin{tabular}{|c|}
    \hline
    1 \\
    \hline
    2 \\
    \hline
    3 \\
    \hline
    4 \\
    \hline
  \end{tabular},
  \begin{tabular}{|c|}
    \hline
    5 \\
    \hline
  \end{tabular},
  \begin{tabular}{|c|}
    \hline
    $\bar{6}$ \\
    \hline
    $\bar{7}$ \\
    \hline
  \end{tabular}\right)_{~FCS}
\end{align}
\begin{itemize}
  \item $I=\frac{1}{2}$ : The wave function is the same as in the case of $q^4 s \bar{Q}^2$ with $I=2$.
\end{itemize}
\subsection{$s^5 \bar{Q}^2$ : \{12345\}\{67\}}
We fix the position of each quarks on $s(1)s(2)s(3)s(4)s(5)\bar{Q}(6)\bar{Q}(7)$. In this case, the flavor $\otimes$ color $\otimes$ spin wave function should satisfy the following symmetry : \{12345\}\{67\}
\begin{align}
  \psi_{FCS}=\left(
  \begin{tabular}{|c|}
    \hline
    1 \\
    \hline
    2 \\
    \hline
    3 \\
    \hline
    4 \\
    \hline
    5 \\
    \hline
  \end{tabular},
  \begin{tabular}{|c|}
    \hline
    $\bar{6}$ \\
    \hline
    $\bar{7}$ \\
    \hline
  \end{tabular}\right)_{~FCS}
\end{align}
\begin{itemize}
  \item $I=0$ : The wave function is the same as in the case of $q^5 \bar{Q}^2$ with $I=\frac{5}{2}$.
\end{itemize}
\begin{center}
\begin{table}[htbp]
  \begin{center}
    \begin{tabular}{|c|c|c|c|c|c|c|c|}
    \hline
     & Isospin & Spin & ~$M$~ &  & Isospin & Spin & ~$M$~ \\
    \hline
    $q^5\bar{Q}^2$ & $\frac{5}{2}$ & $\frac{3}{2}$ & 1 & $q^3s^2\bar{Q}^2$ & $\frac{3}{2}$ & $\frac{7}{2}$ & 1 \\
                   &               & $\frac{1}{2}$ & 1 &                   &               & $\frac{5}{2}$ & 3 \\
    \cline{2-4}
                   & $\frac{3}{2}$ & $\frac{5}{2}$ & 1 &                   &               & $\frac{3}{2}$ & 8 \\
                   &               & $\frac{3}{2}$ & 3 &                   &               & $\frac{1}{2}$ & 7 \\
    \cline{6-8}
                   &               & $\frac{1}{2}$ & 3 &                   & $\frac{1}{2}$ & $\frac{7}{2}$ & 1 \\
    \cline{2-4}
                   & $\frac{1}{2}$ & $\frac{7}{2}$ & 1 &                   &               & $\frac{5}{2}$ & 5 \\
                   &               & $\frac{5}{2}$ & 3 &                   &               & $\frac{3}{2}$ & 13 \\
                   &               & $\frac{3}{2}$ & 4 &                   &               & $\frac{1}{2}$ & 13 \\
    \cline{5-8}
                   &               & $\frac{1}{2}$ & 3 & $q^2s^3\bar{Q}^2$ & 1 & $\frac{7}{2}$ & 1 \\
    \cline{1-4}
    $q^4s\bar{Q}^2$ & 2 & $\frac{5}{2}$ & 1 &                   &   & $\frac{5}{2}$ & 3 \\
                    &   & $\frac{3}{2}$ & 4 &                   &   & $\frac{3}{2}$ & 8 \\
                    &   & $\frac{1}{2}$ & 4 &                   &   & $\frac{1}{2}$ & 7 \\
    \cline{2-4} \cline{6-8}
                    & 1 & $\frac{7}{2}$ & 1 &                   & 0 & $\frac{5}{2}$ & 3 \\
                    &   & $\frac{5}{2}$ & 5 &                   &   & $\frac{3}{2}$ & 6 \\
                    &   & $\frac{3}{2}$ & 10 &                  &   & $\frac{1}{2}$ & 7 \\
    \cline{5-8}
                    &   & $\frac{1}{2}$ & 10 & $qs^4\bar{Q}^2$ & $\frac{1}{2}$ & $\frac{5}{2}$ & 1 \\
    \cline{2-4}
                    & 0 & $\frac{7}{2}$ & 1 &                 &               & $\frac{3}{2}$ & 4 \\
                    &   & $\frac{5}{2}$ & 3 &                 &               & $\frac{1}{2}$ & 4 \\
    \cline{5-8}
                    &   & $\frac{3}{2}$ & 7 & $s^5\bar{Q}^2$ & 0 & $\frac{5}{2}$ & 1 \\
                    &   & $\frac{1}{2}$ & 6 &                &   & $\frac{3}{2}$ & 3 \\
                    &   &               &   &                &   & $\frac{1}{2}$ & 3 \\
    \hline
    \end{tabular}
  \end{center}
  \caption{All the possible heptaquark states containing two heavy antiquarks with the corresponding multiplicity. $M$ represents the multiplicity of the color $\otimes$ flavor $\otimes$ spin state.}
  \label{multiplicity}
\end{table}
\end{center}
We show  all the possible heptaquark state for each flavor, isospin, spin with the corresponding multiplicity in the Table~\ref{multiplicity}.
\section{Color-spin interaction}
\label{Color spin interaction}
In this article, we investigate the stability of the heptaquark configurations using the hyperfine potential given as
\begin{align}
  H=-A\sum_{i<j}\frac{1}{m_i m_j}\lambda_i^c \lambda_j^c \sigma_i \cdot \sigma_j,
  \label{hyper}
\end{align}
where $m_i$'s are the constituent quark masses, and $\lambda^c_i/2$ are the color operator of the $i$'th quark for the color SU(3), and $A$ is taken to be a constant determined from its contribution to the  proton mass using the comprehensive Hamiltonian~\cite{Park:2016mez}. The expectation value of the hyperfine potential of a proton is approximately $-160$ MeV. Since $-\sum \lambda_i^c \lambda_j^c \sigma_i \cdot \sigma_j$ for a proton is $-8$, we extract the value $A/m_u^2=20$ MeV.   While this value depends on the wave function of a multiquark state, we take this value to search for possible stable multiquark configurations that can potentially be stable against strong decays.
In this work, for a given flavor and quantum number of a heptaquark configuration, we calculate the matrix elements for Eq.~(\ref{hyper}) for all possible color spin flavor basis, and then diagonalize the matrix to obtain the configuration with the lowest hyperfine interaction strength.

\section{Results}
\label{Results}
Heptaquark can decay into one baryon and two mesons.  The differences in the confining and coulomb potentials are proportional to the two body color force
$\lambda_i \lambda_j$.  As far as the compact heptaquark, baryon and mesons are taken to occupy the same size, the difference in these energies between the heptaquark and the sum of the baryon and two mesons are negligible.
This is so because if the heptaquark, baryon and mesons have the same size, the confining potential will just be proportional to a common value and the sum of all the two body interaction $\sum \lambda_i \lambda_j$, which are equal to -56/3, -8, -16/3 for the heptaquark, baryon and meson respectively.  The main difference comes from the difference in the color-spin potential.
 Therefore, we define the binding potential of the heptaquark as the difference in the color-spin interaction between the heptaquark and the sum of baryon and two mesons as follows:
\begin{align}
  V_B=H_{\mathrm{heptaquark}}-H_{\mathrm{baryon}}-H_{\mathrm{meson1}}-H_{\mathrm{meson2}}.
\end{align}

To search for possible stable configurations, we plot the binding potential of the heptaquark as a function of the heavy quark mass using a variable $\eta$ defined as follows:
\begin{align}
  \eta=1-\frac{m_u}{m_Q}.
\end{align}
Here, we fix the strange quark mass to 632 MeV, which comes from our previous work~\cite{Park:2016mez}. When the flavor of antiquarks is strange, charm, and bottom, $\eta$ values are approximately 0.46, 0.82, 0.93, respectively. Decay channels that give the lowest potential can change as $\eta$ varies.  Hence, in some figures, there are graphs with turning points that have  sudden change in the slope.

\subsection{$q^5 \bar{Q}^2$ : \{12345\}\{67\}}
\begin{figure}[htbp]
  \begin{center}
         \includegraphics[scale=0.8]{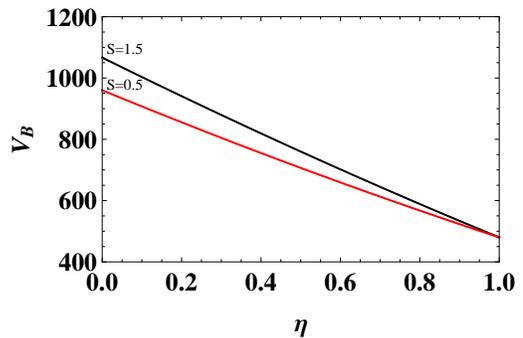}
\caption[]{$V_B$ of $q^5\bar{Q}^2$ with $I=\frac{5}{2}$(unit: MeV).}
\label{fig1}
  \end{center}
\end{figure}

\begin{figure}[htbp]
  \begin{center}
         \includegraphics[scale=0.8]{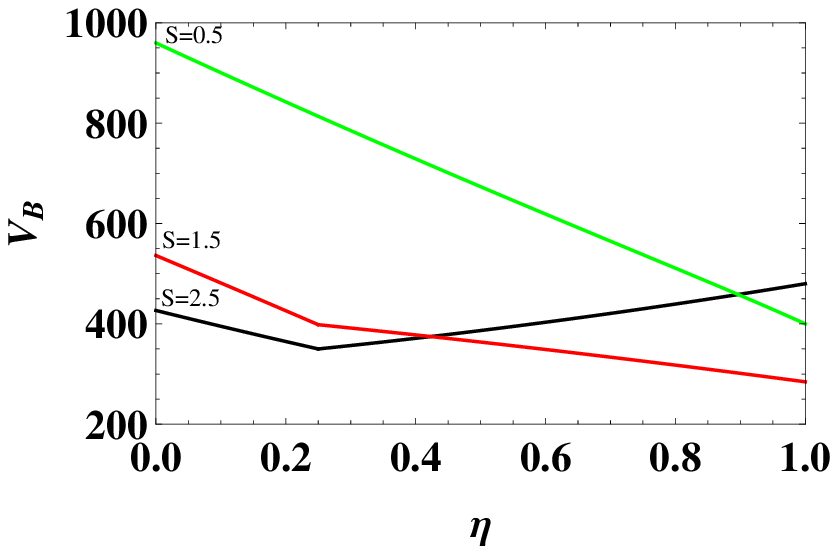}
\caption[]{$V_B$ of $q^5\bar{Q}^2$ with $I=\frac{3}{2}$(unit: MeV).}
\label{fig2}
  \end{center}
\end{figure}

\begin{figure}[htbp]
  \begin{center}
         \includegraphics[scale=0.8]{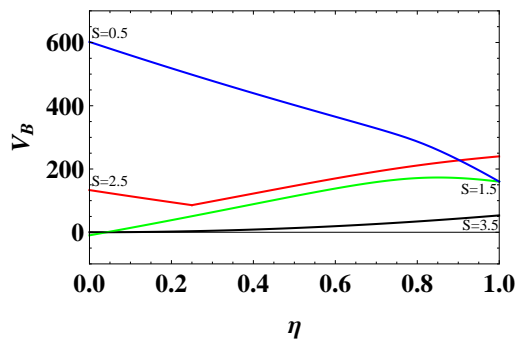}
\caption[]{$V_B$ of $q^5\bar{Q}^2$ with $I=\frac{1}{2}$(unit: MeV).}
\label{fig3}
  \end{center}
\end{figure}

As we can see in Fig~\ref{fig1}-\ref{fig3}, there is no possibility of stable heptaquark except $I=\frac{1}{2}$ and $S=\frac{3}{2}$ when the antiquarks are $\bar{u}$ or $\bar{d}$ quarks. However, the absolute values of the binding potential is very small, so it cannot be compact when we consider the total Hamiltonian including the kinetic term.
\subsection{$q^4 s\bar{Q}^2$ : \{1234\}5\{67\}}
\begin{figure}[htbp]
  \begin{center}
         \includegraphics[scale=0.8]{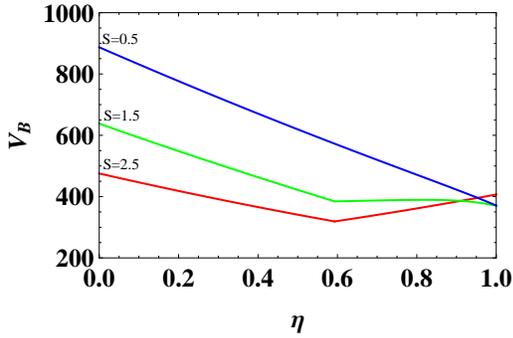}
\caption[]{$V_B$ of $q^4s\bar{Q}^2$ with $I=2$(unit: MeV).}
\label{fig4}
  \end{center}
\end{figure}

\begin{figure}[htbp]
  \begin{center}
         \includegraphics[scale=0.8]{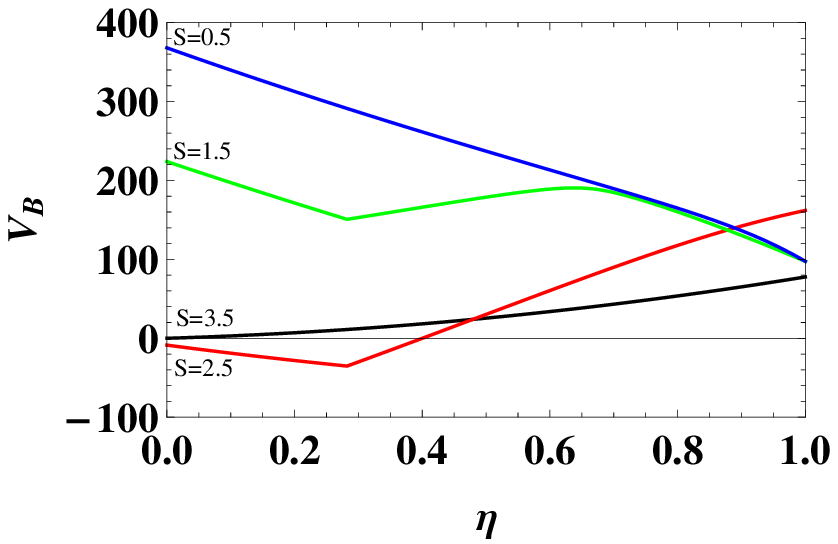}
\caption[]{$V_B$ of $q^4s\bar{Q}^2$ with $I=1$(unit: MeV).}
\label{fig5}
  \end{center}
\end{figure}

\begin{figure}[htbp]
  \begin{center}
         \includegraphics[scale=0.8]{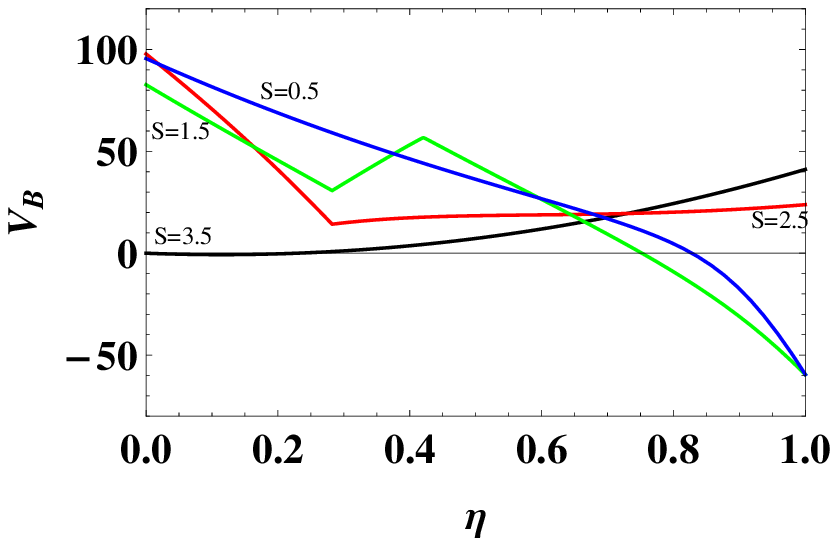}
\caption[]{$V_B$ of $q^4s\bar{Q}^2$ with $I=0$(unit: MeV).}
\label{fig6}
  \end{center}
\end{figure}

In the case of $q^4 s\bar{Q}^2$ with $I=2$ in Fig.~\ref{fig4}, there is no possibility of stable heptaquark. However, in the case with $I=1$ and $S=\frac{5}{2}$ in Fig.~\ref{fig5}, there can be a stable heptaquark when the antiquarks are light quarks. But, the absolute value of the binding potential is still small. In contrast, as can be  seen in Fig.~\ref{fig6}, the heptaquark configuration with $I=0$ and $S=\frac{1}{2}$,$\frac{3}{2}$  can be stable when the mass of antiquarks becomes very large.\\
\subsection{$q^3 s^2 \bar{Q}^2$ : \{123\}\{45\}\{67\}}
\begin{figure}[htbp]
  \begin{center}
         \includegraphics[scale=0.8]{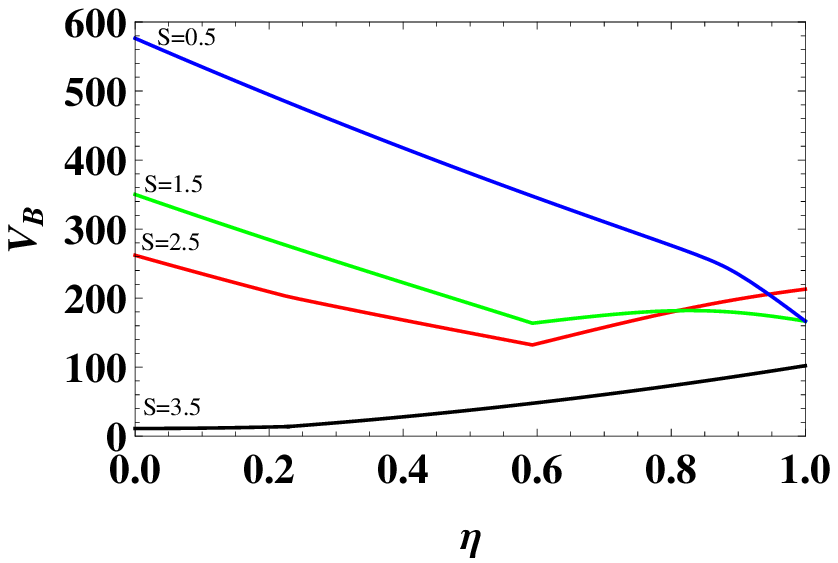}
\caption[]{$V_B$ of $q^3s^2\bar{Q}^2$ with $I=\frac{3}{2}$(unit: MeV).}
\label{fig7}
  \end{center}
\end{figure}

\begin{figure}[htbp]
  \begin{center}
         \includegraphics[scale=0.8]{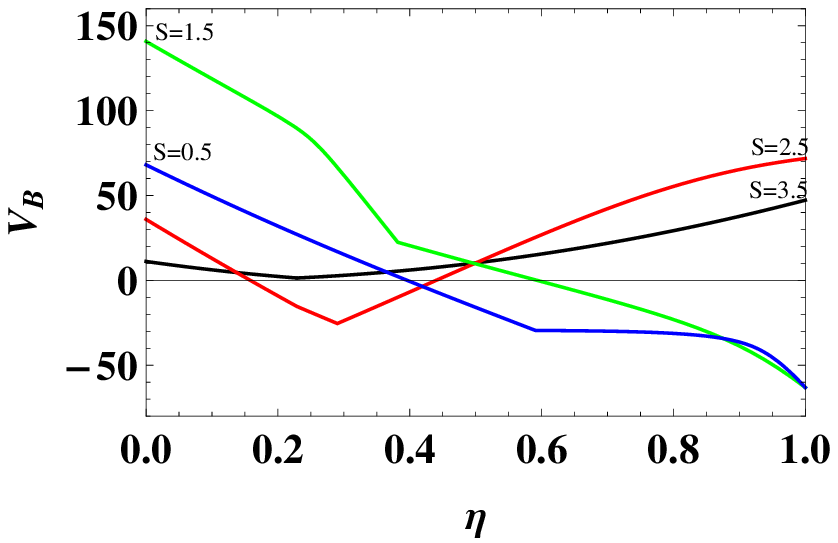}
\caption[]{$V_B$ of $q^3s^2\bar{Q}^2$ with $I=\frac{1}{2}$(unit: MeV).}
\label{fig8}
  \end{center}
\end{figure}

The heptaquark containing two strange quarks with $I=\frac{3}{2}$ in Fig.~\ref{fig7} shows no possibility of a stable heptaquark. In the case of $q^3 s^2 \bar{Q}^2$ with $I=\frac{1}{2}$ and $S=\frac{5}{2}$, as shown in Fig.~\ref{fig8}, there is  a configuration with a slight negative binding potential when the antiquarks are light quarks. Furthermore, for  $I=\frac{1}{2}$ and $S=\frac{1}{2},\frac{3}{2}$ configurations, the potential becomes attractive when  the mass of antiquarks becomes very large.
We represent the expectation values of the hyperfine potential for the heptaquark configuration with $S=\frac{1}{2}$ and the lowest decay mode in Table ~\ref{table-1}. We take the charm quark mass to be  1930 MeV as extracted from fits to the heavy baryon masses using variational method~\cite{Park:2016mez}.
For the $q^3 s^2 \bar{s}^2(I=\frac{1}{2},S=\frac{1}{2})$ case, there is an additional interaction between the $u$ quarks as compared to the isolated baryon  meson states. However, the strength of the interaction between $u$ quark and $s$ quark is reduced. When the antiquarks are heavy quarks, there is an additional repulsion between the $s$ quarks, while there is also an additional attraction $u$ quark and $s$ quark, making the binding potential negative.

As we mentioned in introduction, there is a possibility of a stable heptaquark state as long as there is a stable meson state composed of two heavy quarks and two light quarks within the chiral soliton model~\cite{Bander:1994sp}. It is well known that $T_{cc}$ with $J^P=1^+$, $I=0$ could be a stable tetraquark state\cite{Carlson:1987hh,Lee:2009rt}. Therefore, taking the result in Ref.~\cite{Bander:1994sp} to be valid, there should be a stable configuration composed of five light quarks and two heavy antiquarks with $S=\frac{3}{2}$ or $S=\frac{1}{2}$ and $I=\frac{1}{2}$. It should be noted that although our results for $q^5 \bar{Q}^2$ with $I=\frac{1}{2}$ does not support a stable heptaquark, the configuration with  $q^3 s^2 \bar{Q}^2$ with $S=\frac{3}{2},\frac{1}{2}$ and $I=\frac{1}{2}$ indeed may be stable heptaquark states. As we can see in Table~\ref{multiplicity}, since these two states have large multiplicities compared to the other states, it may lead a low binding potential.
\begin{table}
\caption{The expectation values of the hyperfine potential divided by constant factor $A$ for $q^3 s^2 \bar{Q}^2(I=\frac{1}{   2},S=\frac{1}{2})$ and the corresponding lowest decay mode.}
\begin{center}
    \begin{tabular}{|c|c|}
      \hline \hline
      Heptaquark & The lowest decay mode \\
      \hline
      $q^3 s^2 \bar{s}^2(I=\frac{1}{2},S=\frac{1}{2})$ & $\Xi + K+K$  \\
      $-\frac{2.51}{m_u^2}-\frac{34.22}{m_u m_s}-\frac{5.98}{m_s^2}$ & $-\frac{128}{3m_u m_s}+\frac{8}{3m_s^2}$ \\
      \hline
      $q^3 s^2 \bar{c}^2(I=\frac{1}{2},S=\frac{1}{2})$ & $\Lambda+D+D_s$ \\
      $-\frac{4.3}{m_u^2}-\frac{11.7}{m_u m_s}+\frac{3.01}{m_s^2}$ & $-\frac{8}{m_u^2}-\frac{16}{m_u m_c}-\frac{16}{m_s m_c}$ \\
      $-\frac{15.04}{m_u m_c}-\frac{16.99}{m_s m_c}+\frac{3.07}{m_c^2}$ & \\
      \hline \hline
    \end{tabular}
\end{center}
\label{table-1}
\end{table}
\subsection{$s^3 q^2 \bar{Q}^2$ : \{123\}\{45\}\{67\}}
\begin{figure}[htbp]
  \begin{center}
         \includegraphics[scale=0.8]{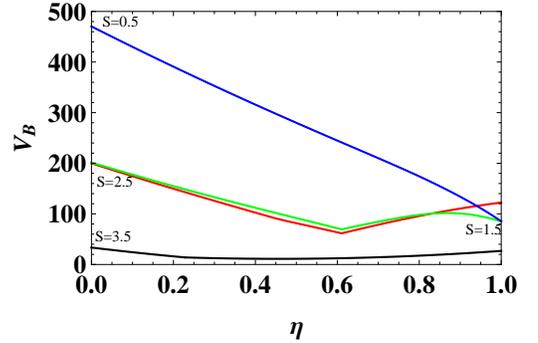}
\caption[]{$V_B$ of $s^3q^2\bar{Q}^2$ with $I=1$(unit: MeV).}
\label{fig9}
  \end{center}
\end{figure}

\begin{figure}[htbp]
  \begin{center}
         \includegraphics[scale=0.8]{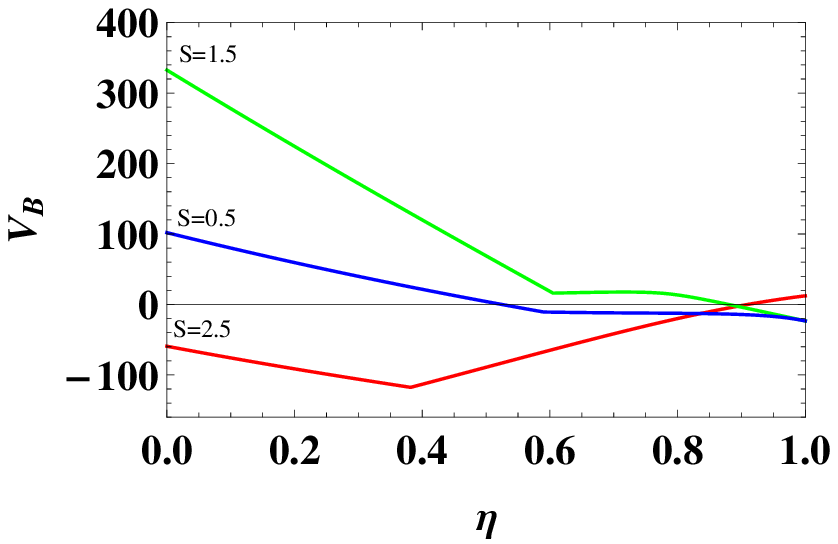}
\caption[]{$V_B$ of $s^3q^2\bar{Q}^2$ with $I=0$(unit: MeV).}
\label{fig10}
  \end{center}
\end{figure}

In this study, the heptaquark with three strange quarks and two antiquarks leads to   the most stable configuration. In the case of $s^3 q^2 \bar{Q}^2$ with $I=1$ in Fig.~\ref{fig9}, there is no possibility of stable heptaquark. As we can see in Fig.~\ref{fig10}, however, there is large negative binding energy with $I=0$ and $S=\frac{5}{2}$. In Table ~\ref{table-2}, we can see there is a considerable amount of additional attraction between  the $u$ and $s$ quarks for $s^3 q^2 \bar{s}^2(I=0,S=\frac{5}{2})$ compared to the corresponding lowest decay mode. However, when the antiquarks are heavy quarks, there is an additional repulsion between $s$ quarks, so it makes the binding potential smaller.

In Table~\ref{additional-kinetic-table}, we present the additional kinetic energy and binding potential of the heptaquark for two most stable cases. Here, we calculate the additional kinetic energy in Eq. (4) with $a_5=a_6= 2.5 \mathrm{fm}^{-2} $, which assumes that the interquark distance of heptaquark is similar to that of a proton. As we can see in the table, when the antiquark is a  heavy quark, the additional kinetic energy is reduced for both cases.

For the $q^3s^2\bar{Q}^2(I=\frac{1}{2},S=\frac{1}{2})$ case, when the antiquarks are heavy quarks, the binding potential is also reduced. However, the additional kinetic energy is still much larger than the absolute value of the binding potential.

For $q^2s^3 \bar{Q}^2(I=0,S=\frac{5}{2})$ case, when the antiquarks are light quarks, the expectation value of the binding potential is largest  and becomes smaller when the antiquarks are heavy quarks.  This is so  because the interaction between $u,d$ quarks and antiquarks is reduced due to the $1/m$ factor. As sizeable repulsion comes from the interaction between the two strange quarks for this quantum number, replacing the strange quark with the heavy quark might lead to a stable heptaquark state.

It should be noted however, that the numbers for the additional kinetic energy shown in Table~\ref{additional-kinetic-table} are obtained assuming that one brings the additional quarks into a compact size of around $\langle r^2 \rangle^{1/2} =a_4^{-1/2} \sim 0.632$ fm.  Assuming that the size becomes larger by a factor of 2, the additional kinetic energy would be reduced by a factor of 4.  Then the $q^3s^2\bar{b}^2(I=\frac{1}{2},S=\frac{1}{2})$ and the $q^2s^3 \bar{s}^2(I=0,S=\frac{5}{2})$ configurations could become stable.  These states will have masses of around 11949 MeV and 3572 MeV, respectively within our model. In particular, the $q^2s^3\bar{s}^2(I=0,S=\frac{5}{2})$ state could decay into $\Lambda+\phi+\phi$, which is easy to reconstruct. The exact value of the mass and the additional kinetic energy  depends on the model employed as we explain the case for MIT bag model in the appendix.  However, the attraction coming from the color-spin interaction will be common to all models and hence the most attractive configurations will point to  the possible stable heptaquark state.

Another interesting possibility is that the string tension in the compact heptaquark configuration will be smaller than those in usual hadrons.  Such possibility has been discussed in Ref.\cite{Buchmann:1998mi} in relation to a stable dibaryon.  The nonperturbative gauge field configuration for generating the confining potential may change in the presence of other color sources and lead to a smaller string tension in a heptaquark or dibaryon configuration. In such cases, even if the heptaquark, baryon and mesons have the same size, the contributions from the confining potential in the heptaquark will be smaller than the sum of the baryon and mesons.  Furthermore, due to a smaller string tension, the wave function of the heptaquark will be more extended leading to a smaller additional kinetic energy.  These two effects  could lead to a more stable and strongly bound heptaquark configuration.    

\begin{table}
\caption{The expectation values of the hyperfine potential divided by constant factor $A$ for $s^3 q^2 \bar{Q}^2(I=0,S=\frac{5}{2})$ and the corresponding lowest decay mode.}
  \begin{center}
    \begin{tabular}{|c|c|}
      \hline \hline
      Heptaquark & The lowest decay mode \\
      \hline
      $s^3 q^2 \bar{s}^2(I=0,S=\frac{5}{2})$ & $\Lambda + \phi + \phi$ \\
      $-\frac{5.97}{m_u^2}-\frac{12.16}{m_u m_s}+\frac{9.28}{m_s^2}$ & $-\frac{8}{m_u^2}+\frac{32}{3m_s^2}$ \\
      \hline
      $s^3 q^2 \bar{c}^2(I=0,S=\frac{5}{2})$ & $\Lambda+D_s^*+D_s^*$ \\
      $-\frac{6.83}{m_u^2}-\frac{4.71}{m_u m_s}+\frac{8.25}{m_s^2}$ & $-\frac{8}{m_u^2}+\frac{32}{3m_s m_c}$ \\
      $-\frac{4.24}{m_u m_c}-\frac{1.32}{m_s m_c}+\frac{2.74}{m_c^2}$ & \\
      \hline \hline
    \end{tabular}
  \end{center}
  \label{table-2}
\end{table}
\begin{table}
\caption{The additional kinetic energy($\Delta K $) and binding potential($V_B$) of the heptaquark. The first table is for $q^3s^2 \bar{Q}^2(I=\frac{1}{2},S=\frac{1}{2})$ and the second one is for $q^2s^3 \bar{Q}^2(I=0,S=\frac{5}{2})$.  In the third table, we represent the parameters used to calculate the additional kinetic energy. The unit of $\Delta K$ and $V_B$ is MeV.}
  \begin{center}
    \begin{tabular}{|c|c|c|c|c|c|c|}
      \hline
      \hline
      $I=\frac{1}{2},S=\frac{1}{2}$ & \multicolumn{2}{|c|}{$q^3s^2 \bar{s}^2$} & \multicolumn{2}{|c|}{$q^3s^2 \bar{c}^2$} & \multicolumn{2}{|c|}{$q^3s^2 \bar{b}^2$} \\
      \hline
      \multirow{2}{*}{$\Delta K $} & $\frac{3\hbar^2}{2M_5}a_5$ & $\frac{3\hbar^2}{2M_6}a_6$ & $\frac{3\hbar^2}{2M_5}a_5$ & $\frac{3\hbar^2}{2M_6}a_6$ & $\frac{3\hbar^2}{2M_5}a_5$ & $\frac{3\hbar^2}{2M_6}a_6$ \\
      \cline{2-7}
       & 388.75 & 294.73 & 210.03 & 139.51 & 163.97 & 65.08 \\
      \hline
      $V_B$ & \multicolumn{2}{|c|}{-9.84} & \multicolumn{2}{|c|}{-31.75} & \multicolumn{2}{|c|}{-40.69} \\
      \hline
      \hline
    \end{tabular}
  \end{center}
  \begin{center}
    \begin{tabular}{|c|c|c|c|c|c|c|}
      \hline
      \hline
      $I=0,S=\frac{5}{2}$ & \multicolumn{2}{|c|}{$q^2s^3 \bar{s}^2$} & \multicolumn{2}{|c|}{$q^2s^3 \bar{c}^2$} & \multicolumn{2}{|c|}{$q^2s^3 \bar{b}^2$} \\
      \hline
      \multirow{2}{*}{$\Delta K $} & $\frac{3\hbar^2}{2M_5}a_5$ & $\frac{3\hbar^2}{2M_6}a_6$ & $\frac{3\hbar^2}{2M_5}a_5$ & $\frac{3\hbar^2}{2M_6}a_6$ & $\frac{3\hbar^2}{2M_5}a_5$ & $\frac{3\hbar^2}{2M_6}a_6$ \\
      \cline{2-7}
       & 271.57 & 245.82 & 201.34 & 135.18 & 162.46 & 63.89 \\
      \hline
      $V_B$ & \multicolumn{2}{|c|}{-98.84} & \multicolumn{2}{|c|}{-16} & \multicolumn{2}{|c|}{3.13} \\
      \hline
      \hline
    \end{tabular}
  \end{center}
  \begin{center}
    \begin{tabular}{|c|c|c|c|c|c|}
      \hline
      \hline
      $m_u$ & $m_s$ & $m_c$ & $m_b$ & $a_5$ & $a_6$ \\
      \hline
      343 & 632 & 1930 & 5305 & 2.5 & 2.5 \\
      MeV & MeV & MeV & MeV & $\mathrm{fm}^{-2}$ & $\mathrm{fm}^{-2}$ \\
      \hline
      \hline
    \end{tabular}
  \end{center}
  \label{additional-kinetic-table}
\end{table}
\subsection{$s^4 q \bar{Q}^2$ : \{1234\}5\{67\}}
\begin{figure}[htbp]
  \begin{center}
         \includegraphics[scale=0.8]{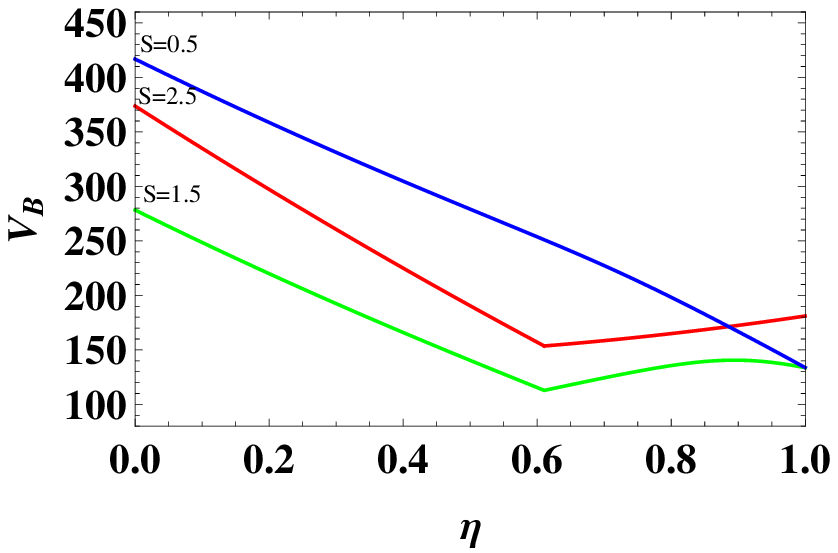}
\caption[]{$V_B$ of $s^4 q\bar{Q}^2$ with $I=\frac{1}{2}$(unit: MeV).}
\label{fig11}
  \end{center}
\end{figure}

The wave function of $s^4 q \bar{Q}^2$ is the same as $q^4 s \bar{Q}^2$ with $I=2$. The only difference is the mass factor in the hyperfine potential. As we can see in Fig.~\ref{fig11}, there is no stable heptaquark with four strange quarks. Additionally, the plot of $s^5 \bar{Q}^2$ is the same as $q^5 \bar{Q}^2$ with $I=\frac{5}{2}$.
\section{summary}
\label{summary}
In this work, we investigated the symmetry property and the stability of the heptaquark containing two identical heavy antiquarks. We constructed the flavor $\otimes$ color $\otimes$ spin wave function satisfying the Pauli principle in the flavor SU(3) breaking case. We then searched for the  heptaquark configuration with the lowest  color-spin interaction, and found that the $s^3 q^2 \bar{s}^2$ with $I=0,S=\frac{5}{2}$ configuration is the most stable state.  For this quantum number, when seven quarks form a compact configuration, the interaction between $u$-$d$ quarks is reduced compared to that in the $\Lambda$, but the  additional interaction between the light quarks and the $s$ quarks result in the additional attracting that could make the heptaquark state stable.  This state could be  probed by reconstructing the $\Lambda+\phi +\phi$ invariant mass or by its weak decay products if it is strongly bound.\\

\section*{acknowledgements}
This work was supported by the Korea National Research Foundation under the grant number 2016R1D1A1B03930089.
The work of W.S. Park was supported  by the National Research Foundation of Korea (NRF) grant funded by the Ministry of Education (No.2017R 1D 1A 1B 03028419).

\appendix
\begin{widetext}
\section{Color basis of the heptaquark}
Here, we present the color basis of the heptaquark using the  tensor form.  The expectation value of all the color operators for the heptaquarks can be obtained using this basis.
\begin{align}
  |C_1 \rangle =& \frac{1}{\sqrt{6}}\{-\frac{\sqrt{3}}{4\sqrt{2}} \varepsilon^{ijk}q^i(1)q^j(2)q^k(3) \varepsilon^{lmn}q^m(4)q^n(5) +\frac{1}{4\sqrt{6}}\varepsilon^{ijk}q^i(1)q^j(2)q^k(4) \varepsilon^{lmn}q^m(3)q^n(5) \nonumber\\
  &-\frac{1}{2\sqrt{6}}\varepsilon^{ijk}q^i(1)q^j(2)q^k(5) \varepsilon^{lmn}q^m(3)q^n(4)-\frac{1}{2\sqrt{6}}\varepsilon^{ijk}q^i(1)q^j(3)q^k(4) \varepsilon^{lmn}q^m(2)q^n(5) \nonumber\\
  &+\frac{1}{\sqrt{6}}\varepsilon^{ijk}q^i(1)q^j(3)q^k(4) \varepsilon^{lmn}q^m(2)q^n(5)\} \varepsilon_{lpr} \bar{q}_p (6) \bar{q}_r (7) \nonumber\\
  |C_2 \rangle =& \frac{1}{\sqrt{6}}\{\frac{1}{4\sqrt{2}} \varepsilon^{ijk}q^i(1)q^j(2)q^k(3) \varepsilon^{lmn}q^m(4)q^n(5) -\frac{1}{4\sqrt{2}}\varepsilon^{ijk}q^i(1)q^j(2)q^k(4) \varepsilon^{lmn}q^m(3)q^n(5) \nonumber\\
  &+\frac{1}{2\sqrt{2}}\varepsilon^{ijk}q^i(1)q^j(2)q^k(5) \varepsilon^{lmn}q^m(3)q^n(4)\} \varepsilon_{lpr} \bar{q}_p (6) \bar{q}_r (7) \nonumber\\
  |C_3 \rangle =& \frac{1}{\sqrt{6}}\{\frac{1}{4\sqrt{2}} \varepsilon^{ijk}q^i(1)q^j(2)q^k(3) \varepsilon^{lmn}q^m(4)q^n(5) -\frac{1}{4\sqrt{2}}\varepsilon^{ijk}q^i(1)q^j(2)q^k(4) \varepsilon^{lmn}q^m(3)q^n(5) \nonumber\\
  &+\frac{1}{2\sqrt{2}}\varepsilon^{ijk}q^i(1)q^j(3)q^k(4) \varepsilon^{lmn}q^m(2)q^n(5)\} \varepsilon_{lpr} \bar{q}_p (6) \bar{q}_r (7) \nonumber\\
  |C_4 \rangle =& \frac{1}{\sqrt{6}}\{\frac{\sqrt{3}}{4\sqrt{2}} \varepsilon^{ijk}q^i(1)q^j(2)q^k(4) \varepsilon^{lmn}q^m(3)q^n(5) -\frac{1}{4\sqrt{6}}\varepsilon^{ijk}q^i(1)q^j(2)q^k(3) \varepsilon^{lmn}q^m(4)q^n(5)\} \varepsilon_{lpr} \bar{q}_p (6) \bar{q}_r (7) \nonumber\\
  |C_5 \rangle =& \frac{1}{\sqrt{6}}\{\frac{1}{2\sqrt{3}} \varepsilon^{ijk}q^i(1)q^j(2)q^k(3) \varepsilon^{lmn}q^m(4)q^n(5)\} \varepsilon_{lpr} \bar{q}_p (6) \bar{q}_r (7) \nonumber\\
  |C_6 \rangle =& \frac{1}{\sqrt{6}}\{\frac{1}{\sqrt{30}} \varepsilon^{ijk}q_i(1)q_j(2)q_k(4) d^{lmn}q^m(3)q^n(5)-\frac{1}{\sqrt{30}} \varepsilon^{ijk}q_i(1)q_j(2)q_k(5) d^{lmn}q^m(3)q^n(4) \nonumber\\
  &+\frac{1}{\sqrt{30}} \varepsilon^{ijk}q_i(1)q_j(3)q_k(4) d^{lmn}q^m(2)q^n(5) -\frac{1}{\sqrt{30}} \varepsilon^{ijk}q_i(1)q_j(3)q_k(5) d^{lmn}q^m(2)q^n(4) \nonumber\\
  &+\frac{\sqrt{3}}{\sqrt{10}} \varepsilon^{ijk}q_i(1)q_j(4)q_k(5) d^{lmn}q^m(2)q^n(3) \} d_{lpr} \bar{q}_p (6) \bar{q}_r (7) \nonumber\\
  |C_7 \rangle =& \frac{1}{\sqrt{6}}\{\frac{\sqrt{3}}{4\sqrt{5}} \varepsilon^{ijk}q_i(1)q_j(2)q_k(3) d^{lmn}q^m(4)q^n(5)+\frac{1}{4\sqrt{15}} \varepsilon^{ijk}q_i(1)q_j(2)q_k(4) d^{lmn}q^m(3)q^n(5) \nonumber\\
  &-\frac{1}{\sqrt{15}} \varepsilon^{ijk}q_i(1)q_j(2)q_k(5) d^{lmn}q^m(3)q^n(4) -\frac{1}{2\sqrt{15}} \varepsilon^{ijk}q_i(1)q_j(3)q_k(4) d^{lmn}q^m(2)q^n(5) \nonumber\\
  &+\frac{2}{\sqrt{15}} \varepsilon^{ijk}q_i(1)q_j(3)q_k(5) d^{lmn}q^m(2)q^n(4) \} d_{lpr} \bar{q}_p (6) \bar{q}_r (7) \nonumber\\
  |C_8 \rangle =& \frac{1}{\sqrt{6}}\{-\frac{1}{4\sqrt{5}} \varepsilon^{ijk}q_i(1)q_j(2)q_k(3) d^{lmn}q^m(4)q^n(5)-\frac{1}{4\sqrt{5}} \varepsilon^{ijk}q_i(1)q_j(2)q_k(4) d^{lmn}q^m(3)q^n(5) \nonumber\\
  &+\frac{1}{\sqrt{5}} \varepsilon^{ijk}q_i(1)q_j(2)q_k(5) d^{lmn}q^m(3)q^n(4) \} d_{lpr} \bar{q}_p (6) \bar{q}_r (7) \nonumber\\
  |C_9 \rangle =& \frac{1}{\sqrt{6}}\{\frac{1}{4} \varepsilon^{ijk}q_i(1)q_j(2)q_k(3) d^{lmn}q^m(4)q^n(5)-\frac{1}{4} \varepsilon^{ijk}q_i(1)q_j(2)q_k(4) d^{lmn}q^m(3)q^n(5) \nonumber\\
  &+\frac{1}{2} \varepsilon^{ijk}q_i(1)q_j(3)q_k(4) d^{lmn}q^m(2)q^n(5) \} d_{lpr} \bar{q}_p (6) \bar{q}_r (7) \nonumber\\
  |C_{10} \rangle =& \frac{1}{\sqrt{6}}\{-\frac{1}{4\sqrt{3}} \varepsilon^{ijk}q_i(1)q_j(2)q_k(3) d^{lmn}q^m(4)q^n(5)+\frac{\sqrt{3}}{4} \varepsilon^{ijk}q_i(1)q_j(2)q_k(4) d^{lmn}q^m(3)q^n(5)\} d_{lpr} \bar{q}_p (6) \bar{q}_r (7) \nonumber\\
  |C_{11} \rangle =& \frac{1}{\sqrt{6}}\{\frac{1}{\sqrt{6}} \varepsilon^{ijk}q_i(1)q_j(2)q_k(3) d^{lmn}q^m(4)q^n(5)\} d_{lpr} \bar{q}_p (6) \bar{q}_r (7)
\end{align}
where the non-vanishing $d^{abc}$ and $d_{abc}$ constants are
\begin{align}
  &d^{111}=d_{111}=d^{222}=d_{222}=d^{333}=d_{333}=1 \nonumber\\
  &d^{412}=d_{412}=d^{421}=d_{421}=d^{523}=d_{523}=d^{532}=d_{532}=d^{613}=d_{613}=d^{631}=d_{631}=\frac{1}{\sqrt{2}}.
\end{align}
\end{widetext}

\section{Kinetic energy}

Consider the simple MIT bag model mass formula for a Hadron composed of $N=N_1+N_2$ quarks\cite{Chodos:1974je,Chodos:1974pn,DeGrand:1975cf} in S wave.
\begin{eqnarray}
E_N=N\frac{\omega}{R}+B\frac{4}{3}\pi R^3-\frac{Z_0}{R}.
\label{bag}
\end{eqnarray}
Here, $\omega\sim 2.04$ and $R,B$ is the bag radius and pressure, respectively.  The last term was originally introduced as the zero point energy or Casimir energy effect but is understood to be taking care of the center of mass motion of the hadrons composed of $N$ quarks\cite{Donoghue:1979ax}.   If the hypothetical hadron decays into two color singlet hadrons of $N_1$ and $N_2$ quarks respectively, their masses will also follow the same formula as Eq.~(\ref{bag}) after replacing the number of quarks to either $N_1$ or $N_2$.  For each hadrons, the bag radius $R$ is determined by minimizing the mass with respect to $R$.  However, comparing the mass of the multiquark composed of $N$ quarks to the sum of two hadrons, one notices that the multiquark state has one less factor of the center of mass term.  This difference is the additional kinetic energy needed to bring the $N_1+N_2$ quarks into a compact configuration compared to two isolated hadrons.

\end{document}